\documentclass[a4paper,11pt]{article}

\usepackage{amsmath,amsfonts,amssymb,caption,graphicx}
\usepackage{ushort}
\usepackage{makeidx}
\usepackage{float}
\usepackage{accents}
\usepackage{color}
\usepackage{xcolor}
\usepackage{framed}
\usepackage{versions}
\usepackage{emptypage}
\usepackage{wrapfig}
\usepackage[T1]{fontenc}
\usepackage[utf8]{inputenc}
\usepackage{titlesec}
\usepackage{fancyhdr}
\usepackage{extramarks}
\usepackage[bookmarks]{hyperref}
\usepackage{hyperref}
\usepackage{bbm}
\usepackage{enumitem}

\usepackage{ragged2e}
\hypersetup{
    colorlinks=true,   	% false: boxed links; true: colored links
    linkcolor=red,      % color of internal links (change box color 
    citecolor = [rgb]{0 0.7 0},   	% color of links to bibliography
    filecolor=magenta, 	% color of file links
    urlcolor=blue
}

\newcommand{\Xp}{\mbox{\boldmath $X$}}

\newcommand{\RL}{{\mathbb R}}

\newcommand{\IND}{{\mathbb I}}
\newcommand{\BBP}{{\mathbb P}}

\newcommand{\VAR}{\mbox{\rm Var}}

\newcommand{\Bern}{\mbox{\rm Bern}}

\def\ba{\begin{align}}
\def\ea{\end{align}}
\def\ban{\begin{align*}}
\def\ean{\end{align*}}

\def\be{\begin{eqnarray}}
\def\ee{\end{eqnarray}}
\def\ben{\begin{eqnarray*}}
\def\een{\end{eqnarray*}}

\def\bqq{\begin{equation}}
\def\eqq{\end{equation}}
\def\bqqn{\begin{equation*}}
\def\eqqn{\end{equation*}}

\def\sq{$\Box$}

\def\qed{\ifmmode\sq\else{\unskip\nobreak\hfil
\penalty50\hskip1em\null\nobreak\hfil\sq
\parfillskip=0pt\finalhyphendemerits=0\endgraf}\fi\par\medbreak}

\newsavebox{\junk}
\savebox{\junk}[1.6mm]{\hbox{$|\!|\!|$}}

\def\til={{\widetilde =}}

\def\clA{{\cal A}}

\def\clP{{\cal P}}

\def\clR{{\cal R}}
\def\clS{{\cal S}}
\def\clT{{\cal T}}

 \def\eq#1/{(\ref{#1})}

\newtheorem{theorem}{Theorem}[section]

\newtheorem{lemma}[theorem]{Lemma}

\def\eq#1/{(\ref{e:#1})}

\def\bdes{\begin{description}}
\def\edes{\end{description}}

\def\notes#1{}

\definecolor{mag}{rgb}{0.7,0,0.3}
\definecolor{dgreen}{rgb}{0.1,0.5,0.1}
\definecolor{dred}{rgb}{.8,0,0}
\definecolor{gray}{rgb}{.8,.8,.8}
\definecolor{brown}{rgb}{0.6451,0.3706,0.1745}

\newcommand{\R}{\mathbb{R}}

\title{Pragmatic lossless compression:\\ 
Fundamental limits and universality}

\author{
Andreas Theocharous
\and 
Lampros Gavalakis
\and 
Ioannis Kontoyiannis
}

\date{\today}

\begin{document}

\maketitle

\footnotetext{All three authors are with the
Statistical Laboratory, Centre for Mathematical Sciences, 
University of Cambridge, Wilberforce Road, Cambridge CB3 0WB, 
UK. Email: \texttt{at771@cam.ac.uk},
\texttt{lg560@cam.ac.uk},
\texttt{yiannis@maths.cam.ac.uk}.\\
This work was supported in part
by the EPSRC-funded INFORMED-AI project EP/Y028732/1.\\
A preliminary version of some of the results in this paper
appeared at ISIT 2025~\cite{theocharous:ISIT25}.}

\begin{abstract}
The problem of variable-rate lossless data compression
is considered,
for codes with and without prefix constraints.
 Sharp bounds are derived for the best 
achievable compression rate of memoryless sources,
when the excess-rate probability is required to be
exponentially small in the blocklength.
Accurate nonasymptotic expansions 
with explicit constants are obtained
for the optimal rate, using tools from large deviations 
and Gaussian approximation. When the source
distribution is unknown, a universal achievability
result is obtained with an explicit ``price for 
universality'' term. This is based on a fine
combinatorial estimate on the number of sequences
with small empirical entropy, which might be
of independent interest.
Examples are shown indicating that, in the 
small excess-rate-probability 
regime, the approximation to the fundamental limit
of the compression rate suggested by these
bounds is significantly more accurate
than the approximations provided by either 
normal approximation or error exponents.
The new bounds reinforce the crucial 
operational conclusion that,
in applications where the blocklength is relatively
short and where stringent guarantees are required
on the rate, the best achievable rate is no longer 
close to the entropy. Rather, it is an appropriate,
more {\em pragmatic} rate, determined via the inverse 
error exponent function and the blocklength.

% THIS PAPER IS ELIGIBLE FOR THE STUDENT PAPER AWARD.
\end{abstract}


\section{Introduction}
%%%%%%%%%%%%%%%%%%%%%%%%%%%%%%%%%%%%%%%%%%%%%%%%%%%%%%%%%%%%%%%%%%%%%%%%%%%%%%%
\subsection{Variable-rate data compression}
%%%%%%%%%%%%%%%%%%%%%%%%%%%%%%%%%%%%%%%%%%%%%%%%%%%%%%%%%%%%%%%%%%%%%%%%%%%%%%%
\label{s:VRDC}

Let $\Xp=\{X_n\;;\;n\geq 1\}$ be a memoryless source consisting
of independent and identically distributed (i.i.d.) random variables
$X_n$, with values in a finite alphabet $A$, and
with common distribution described by the probability 
mass function (p.m.f.) $P$:
$\BBP(X_n=x)=P(x)$, $x\in A$.

We revisit the problem of losslessly compressing the 
output of such a memoryless source, when its distribution
is known {\em a priori} as well as when it is not.
The problem formulation is
simple and quite elementary, though its importance
can hardly be overstated, in view of its application
across the sciences and engineering.

For strings of symbols from $A$, we write $x^n$ for
the block $x^n=(x_1,x_2,\ldots,x_n)$, and similarly,
for blocks of random variables, $X^n$ denotes
$(X_1,X_2,\ldots,X_n)$.
A variable-rate code, or
{\em fixed-to-variable compressor}, for strings of length $n$ 
from $A$ is an injective function 
$f_n: A^n \to \{ 0,1\}^*$, where $\{0,1\}^*$ denotes the set of all
finite-length binary strings:
\begin{equation*}
   \{ 0,1\}^* = \{ \varnothing, 0, 1, 00, 01, 10, 11, 000 ,001 , \ldots \}.    
\end{equation*}
Writing $\ell(s)$ for the length of a string $s\in \{ 0,1\}^*$,
the compressor $f_n$ maps strings 
$x^n \in A^n$ into binary strings $f_n(x^n)$ of length $\ell(f_n(x^n))$ bits. 
We call such compressors {\em one-to-one codes}.

We are interested in the 
best achievable performance among all 
fixed-to-variable compressors.
For a blocklength $n\geq 1$ and rate $R>0$,
the best achievable excess-rate probability 
$\epsilon_n^*(R,P)$ for a source $\Xp$ 
with distribution $P$
is:
\begin{equation*}
    \epsilon_n^*(R,P):= \min_{f_n} 
\mathbb{P}\big(\ell(f_n(X^n)) \geq nR\big). 
\end{equation*}
This minimum is achieved by an optimal compressor
$f_n^*$ independent of the rate $R$~\cite{kontoyiannis-verdu:14}.
Similarly, for a blocklength $n\geq 1$ and excess-rate probability
$\epsilon>0$, the best achievable rate
$R_n^*(\epsilon,P)$, is:
$$R_n^*(\epsilon,P) 
:= 
	\inf\big\{ R>0 : \min_{f_n}\mathbb{P}\big(\ell(f_n(X^n)) \geq nR\big)
	\leq \epsilon\big\}.
$$

\noindent
{\bf Prefix-free codes.}
In the case of prefix-free codes,
analogous fundamental limits,
$\epsilon^{\sf p}_n(R,P)$ and $R_n^{\sf p}(\epsilon,P)$,
can be defined,
by restricting the corresponding minimisations to 
codes that satisfy the prefix condition.
But $R_n^{\sf p}(\epsilon,P)$ is very 
tightly coupled with $R_n^*(\epsilon,P)$: As
shown in \cite[Theorem~1]{kontoyiannis-verdu:14}, 
we always have $R^{\sf p}_n(\epsilon,P)=R_n^*(\epsilon,P)+\frac{1}{n}$.

\subsection{Background}
%%%%%%%%%%%%%%%%%%%%%%%%%%%%%%%%%%%%%%%%%%%%%%%%%%%%%%%%%%%%%%%%%%%%
\label{s:back}

When the source distribution $P$ is known,
and the excess-rate probability $\epsilon$ is fixed,
the problem of determining the first-order asymptotic 
behaviour of the optimal rate $R_n^*(\epsilon,P)$ essentially
goes back to Shannon. We have,
\bqq
nR_n^*(\epsilon,P)= nH(P)+o(n)\;\;\mbox{bits},
\quad\mbox{as}\;n\to\infty,
\label{eq:shannon1}
\eqq
where $H(P)=-\sum_x P(x)\log P(x)$
as usual
denotes the entropy, 
and, throughout, `log' denotes
$\log_2$, 
the logarithm taken to base~2;
see~\cite{kontoyiannis-verdu:14} for an extensive
discussion and historical comments.

The first-order relation~(\ref{eq:shannon1})
was refined by Strassen~\cite{strassen:64b} who
claimed that, as $n\to\infty$,
\bqq
nR_n^*(\epsilon,P) = nH(P) +\sqrt{n}\sigma(P)Q^{-1}(\epsilon) 
- \frac{1}{2}\log n  + O(1)\;\;\mbox{bits}, 
\label{eq:strassen}
\eqq
where $\sigma^2(P)=\VAR_P(-\log P(X))$ is the 
{\em minimal coding variance}~\cite{kontoyiannis-97} or
{\em source dispersion}~\cite{kontoyiannis-verdu:14},
and $Q(x)=1-\Phi(x)$,
% = \int_x^\infty \frac{1}{\sqrt{2\pi}}e^{-z^2/2}\,dz$,
$x\in\RL$, denotes the
standard normal tail function.
In fact, Strassen identified the $O(1)$ term explicitly,
giving an expansion with an $o(1)$ error term.
Although some issues of rigour were raised 
in~\cite{kontoyiannis-verdu:14}
regarding his proof, a stronger
version of~(\ref{eq:strassen}) was established 
in~\cite{kontoyiannis-verdu:14},
where explicit, finite-$n$ bounds were obtained for the $O(1)$
term.

When the excess-rate probability is not fixed but is 
instead required to decay to zero exponentially fast, 
the best rate that can be achieved turns out to be
higher than the entropy. Specifically,
for $\delta>0$ in an appropriate range, as $n\to\infty$,
\bqq
nR_n^*(2^{-n\delta},P)=nH(P_{\alpha^*})+o(n)\;\;\mbox{bits},
\label{eq:blahut}
\eqq
for a specific $\alpha^*\in(0,1)$ depending on $\delta$,
where, for each $\alpha\in(0,1)$, $P_\alpha$ denotes
the tilted distribution,
$$P_{\alpha}(x)=\frac{P(x)^\alpha}{\sum_{y\in A}P(y)^{\alpha}},
\quad x\in A.$$
The first-order asymptotic relation~(\ref{eq:blahut}) is 
simply the ``rate version'' of the well-known
error-exponents result that was 
established in the early works~\cite{dobrushin:62,jelinek:book}
and in the present form by Blahut~\cite{blahut:74}.
Indeed, $H(P_{\alpha^*})$ can be expressed as the 
inverse of the error-exponent function,
$$\Delta_P(R):=\inf_{P':H(P')\geq R} D(P'\|P),$$
via
$$H(P_{\alpha^*})=\Delta_P^{-1}(\delta)=\sup_{P':D(P'\|P)\leq \delta} H(P'),$$
where $D(P'\|P)$ is the relative entropy between $P'$ and $P$.

\subsection{Compression at pragmatic rates: Known source}
%%%%%%%%%%%%%%%%%%%%%%%%%%%%%%%%%%%%%%%%%%%%%%%%%%%%%%%%%%%%%%

The first main contribution of the present work is 
a nonasymptotic refinement of the evaluation
of the fundamental compression limit $R_n^*$, in
the small excess-rate probability regime,
when the source distribution $P$ is known and fixed.
In the same sense in which Strassen's
expansion~(\ref{eq:strassen}) and 
the finite-$n$ bounds of~\cite{kontoyiannis-verdu:14}
strengthen the classical Shannon asymptotic 
relation~(\ref{eq:shannon1}),
we show it is possible to derive 
{\em finite-$n$} bounds that provide 
a correspondingly finer and stronger version of the 
first-order asymptotic 
expansion~(\ref{eq:blahut}) in
the small excess-rate probability regime.

In Theorems~\ref{thm_achievability} and~\ref{thm_converse}
of Section~\ref{s:main} we provide explicit constants
$C$ and $C'$ such that,
\bqq
C'\leq
nR_n^*(2^{-n\delta},P) -\Big[ nH(P_{\alpha^*})
-\frac{1}{2(1-\alpha^*)}\log n\Big]
	\leq C,
\label{eq:main}
\eqq
for all $n$ greater 
than some explicit $N_0$. In view of this,
we may call the expression
$$\clR_n(\epsilon,P):=H(P_{\alpha^*})
-\frac{1}{2n(1-\alpha^*)}\log n\;\;\mbox{bits/symbol},$$
the best {\em pragmatic rate} that can be
achieved at blocklength $n$ with excess-rate
probability no greater than $\epsilon=2^{-n\delta}$.

On account of the remark at the end of Section~\ref{s:VRDC},
the exact same bounds as in~(\ref{eq:main}) hold in the
case of codes with prefix constraints  -- that is, 
for $R_n^{\sf p}$ in place of $R_n^*$ -- 
with the same constant $C'$, and with $C+1$ in place of $C$.

Despite their technical nature,
the approximations to $R_n^*$ and $R_n^{\sf p}$
provided by all these different
approaches are of as much practical relevance as they 
are of mathematical interest. In particular, they each
are useful in different regimes of the blocklength
and the excess-rate probability requirements.
For example, the Strassen-style approximation
suggested by~(\ref{eq:strassen})
is only relevant for moderate values of $\epsilon$. 
If $\epsilon$ is small, then the second term dominates
and the approximation is no longer valid or useful.

In that case, the operational utility of~(\ref{eq:main}) can be
described as follows. Given the blocklength~$n$
and a small target excess-rate probability $\epsilon$,
we can compute $\delta=(1/n)\log(1/\epsilon)$ and
solve for the corresponding $\alpha^*$. Then
the best achievable rate with these parameters
is approximately $\clR_n(\epsilon,P)$ bits/symbol.
As a simple, concrete example, 
consider the case of binary memoryless source $\Xp=\{X_n\}$
where each $X_n$ has p.m.f.\
$P \sim \Bern(0.2)$.

We examine the four approximations to $R_{n}^*(\epsilon,P)$
suggested by the above results
for various small values of $\epsilon$.
In the `Shannon regime',
\begin{align}
\mbox{Shannon:}\qquad &R_n^*\approx H(P),
	\label{eq:stein}\\
\mbox{Strassen:}\qquad &R_n^*\approx H(P)
	+\frac{1}{\sqrt{n}}\sigma(P) Q^{-1}(\epsilon)-\frac{1}{2n}\log n,
	\label{eq:strassen2}
\end{align}
and in the `error exponents regime',
\begin{align}
\mbox{Blahut:}
	\qquad &R_{n}^*\approx H(P_{\alpha^*}),
	\label{eq:hoeffding}\\
\mbox{Thms.~\ref{thm_achievability}-\ref{thm_converse}:}
	\qquad &R_{n}^*\approx \clR_n=H(P_{\alpha^*})
	-\frac{1}{2n(1-\alpha^*)}\log n.
	\label{eq:newapprox}
\end{align}
Table~\ref{Table:n=50} shows 
representative results 
when the blocklength $n = 50$,
which clearly demonstrate that the
pragmatic rate $\clR_n(\epsilon,P)$ provides by far the 
most accurate estimate
of the optimal rate $R_{n}^*(\epsilon,P)$
for the problem parameters considered.

\begin{table}[ht!]
\centering
\begin{tabular}{|c||c|c|c|c|c|}
\hline
	\multicolumn{1}{|c||}{$\epsilon$} 
	& \multicolumn{1}{|c|}{$R_{n}^*(\epsilon,P)$}
	& \multicolumn{1}{c|}{Shannon~(\ref{eq:stein})} 
	& \multicolumn{1}{c|}{Strassen~(\ref{eq:strassen2})}
	& \multicolumn{1}{c|}{Blahut~(\ref{eq:hoeffding})} 
	& \multicolumn{1}{c|}
		% {Thms.~\ref{thm_achievability}-\ref{thm_converse}:~
		{$\clR_n(\epsilon,P)$~(\ref{eq:newapprox})}\\
	\hline
 0.00003 & {\bf 0.940} & 0.722     & 1.119  & 1.000 & {\bf 0.941} \\
 0.00010 & {\bf 0.940} & 0.722     & 1.086  & 0.997 & {\bf 0.936} \\
 0.00032 & {\bf 0.920} & 0.722     & 1.052  & 0.993 & {\bf 0.928} \\
 0.00093 & {\bf 0.900} & 0.722     & 1.017  & 0.987 & {\bf 0.917} \\ 
 0.00251 & {\bf 0.900} & 0.722     & 0.983  & 0.979 & {\bf 0.903} \\ 
 0.00626 & {\bf 0.880} & 0.722     & 0.948  & 0.969 & {\bf 0.888} \\ 
 0.01444 & {\bf 0.840} & 0.722     & 0.913  & 0.957 & {\bf 0.869} \\ 
\hline
\end{tabular}
\centering
\caption{Comparison between the true value of the optimal 
rate $R_{n}^*(\epsilon,P)$ and four different approximations.
Clearly the approximation suggested by Theorems~\ref{thm_achievability}
and~\ref{thm_converse}, given by the pragmatic rate $\clR_n$, gives the best 
results in this regime.}
\label{Table:n=50}
\end{table}

\subsection{Universal compression at pragmatic rates}
%%%%%%%%%%%%%%%%%%%%%%%%%%%%%%%%%%%%%%%%%%%%%%%%%%%%%%%%%%%%%%
\label{s:Iuniversal}

Next, we examine the best rate that can be {\em universally}
achieved, by a single one-to-one compressor applied to 
any memoryless source $\Xp$ on a finite alphabet $A$ of size $m$.

In the Shannon regime,
Kosut and Sankar~\cite{kosut:17} showed that
the same optimal performance as for known
sources~\cite{strassen:64b,kontoyiannis-verdu:14}
could be achieved universally by one-to-one codes, 
at a cost of approximately 
$$\Big(\frac{m-2}{2}\Big)\log n \quad\mbox{bits},$$
over all memoryless sources on an alphabet 
of size $m=|A|\geq 2$.
In Theorem~\ref{thm:universal} in Section~\ref{s:universal},
we prove an analogous result in the error exponents
regime. Namely, we describe a specific sequence
of one-to-one compressors $\{\phi_n^*\}$ that
simultaneously achieve excess-rate probability no greater
than $2^{-n\delta}$ on any memoryless source with
full support on $A$, at a rate no greater than:
\bqq
H(P_{\alpha^*}) 
+ \Big(\frac{m-2}{2} - \frac{1}{2(1-\alpha^* )} \Big) \frac{\log n}{n}
+ O\Big(\frac{1}{n}\Big)\quad\mbox{bits per symbol}.
\label{eq:unirate}
\eqq
Interestingly, the code used to prove Theorem~\ref{thm:universal}
is somewhat simpler than the compressors employed in~\cite{kosut:17}.
The idea is to simply order all string $x^n\in A^n$ in order of 
increasing empirical entropy, and take $\phi_n^*(x^n)$ to be
the binary description of the index of $x^n$ in this ordering.

On the other hand, in the present setting
the analysis of their performance 
requires a very accurate
combinatorial estimate on the number of strings with low
empirical entropy. Specifically, in Theorem~\ref{thm:Theta}
we show that the number of $x^n\in A^n$ such that
$H(\hat{P}_{x^n})\leq H(Q)$ is:
$$\Theta\big(n^{\frac{m-3}{2}}2^{nH(Q)}\big).$$
The proof of this estimate is the most technical
part of this paper, and the result itself may be of
independent interest.

Although we do not provide a matching converse
to the achievability statement in Theorem~\ref{thm:universal}, 
we conjecture that the rate obtained there,
given in~(\ref{eq:unirate}), is optimal.

At first glance, it may be surprising that 
the cost of universality, both in the
Shannon regime and in the error exponents regime,
is $(\frac{m-2}{2})\log n$ bits, instead
of the more familiar 
$(\frac{m-1}{2})\log n$ bits seen in the 
classical universal compression literature
on compression redundancy.
This discrepancy comes from the fact
that, unlike, e.g., in Shtarkov's~\cite{shtarkov:77}
or Rissanen's~\cite{rissanen:84} well-known results,
we consider one-to-one codes instead of prefix-free
codes. In the case of a known source, 
as describer earlier, this makes a difference
of at most one bit~\cite{kontoyiannis-verdu:14}.
But in the case of universal compression,
the difference between optimal prefix-free 
and optimal one-to-one
codes has been found to be of the order of $\log n$ bits,
both in the Shannon excess-rate regime~\cite{kosut:17},
and in terms of redundancy rates~\cite{beirami:14}.

One way to see why there is, indeed, a difference
in universal coding with one-to-one versus with
prefix-free codes,
is to consider binary memoryless sources.
As described in~\cite{kontoyiannis-verdu:14},
the optimal one-to-one code of any such source simply
orders all strings of length $n$ in decreasing
probability, and assigns
binary descriptions of increasing lengths 
to their indices lexicographically.
But for binary ${\rm Bern}(p)$ sources there are only
two optimal orderings, depending on whether~$p$ 
is greater or smaller than~1/2. Therefore,
a universal one-to-one code only needs one extra bit to 
differentiate between these to cases. On the other hand,
the optimal prefix-free universal code needs to describe
the maximum likelihood estimate of $p$ based on $x^n$,
to $\frac{1}{\sqrt{n}}$ accuracy, which takes about
$\frac{1}{2}\log n$ bits; see, e.g., the discussions
in~\cite{barron:02} or~\cite{grunwald:book}. So there is
an $O(\log n)$ universality penalty for prefix-free
codes, whereas there is only an $O(1)$ universality
penalty for one-to-one codes.

We do not consider the problem of 
prefix-free universal codes in the error exponents regime in this
work, but we conjecture that, compared to the optimal
rate in~\eqref{eq:main}, they incur a penalty of,
at best, $\frac{m}{2}\log n$ bits; cf.~\cite{kosut:17}.

\subsection{Prior related work}
%%%%%%%%%%%%%%%%%%%%%%%%%%%%%%%%%%%%%%%%%%%%%%%%%%%%%%%%%%%%%%

Formally, the problem of lossless data compression is very
closely related to binary hypothesis testing, especially in the
case of fixed-rate compression. Indeed, there are close 
hypothesis-testing parallels 
to all the major data-compression
steps outlined in Section~\ref{s:back},
beginning
with Stein's lemma~\cite{chernoff:52,kullback-book,cover:book2},
its refinement by Strassen~\cite{strassen:64b} and 
Polyanskiy-Poor-Verd\'{u}~\cite{PPV:10}, 
and the error-exponent asymptotics of
Hoeffding~\cite{hoeffding:65}. 

For hypothesis testing, sharper expansions in the error-exponents
regime have been developed by numerous authors, 
including the work reported in~\cite{csiszar:71,altug:11,%
altug:14,vazquez:18,nakibouglu:19,nakibouglu:20}.
The results in this paper can be viewed as lossless 
compression analogues of the refined error-exponent bounds 
obtained for binary hypothesis testing
in~\cite{nakibouglu:19,nakibouglu:20,lungu-K:ISIT24,lungu-arxiv:24}.

For fixed-rate data compression,
Csisz{\'a}r and Longo~\cite{csiszar:71}
obtained an asymptotic expansion for
the excess-rate probability
which is accurate up to and including the $O(1)$ term, 
when the target rate is fixed. 
B.~Nakibo{\u{g}}lu~\cite{nakibouglu:19}
observed that this $O(1)$ term is
incorrect in~\cite{csiszar:71},
and there are also further questions
of rigour regarding the proofs there,
partly stemming from the fact that they
rely on earlier results of Strassen~\cite{strassen:64b}.
Our Theorems~\ref{thm_achievability}
and~\ref{thm_converse} can be viewed as finite-$n$
improvements of the ``rate version'' of these expansions,
in the variable-rate case.
Also, in a related but different direction, the best achievable
rate in the ``moderate deviations'' regime was examined
in~\cite{altug-wagner-K:13}. 

Extensive discussions 
of different aspects of the fundamental limits of lossless
data compression 
can be found in Csisz\'{a}r and K\"{o}rner's classic
text,~\cite{csiszar:book2},
Han's book on information spectrum methods~\cite{han:book-en},
and Tan's monograph~\cite{tan:book}.
Universal lossless data compression also has a rich history. The 
texts~\cite{rissanen:book,csiszar-shields:04,grunwald:book,cover:book2}
and the papers~\cite{rissanen:84,barron-rissanen-yu:98}
provide comprehensive overviews.

\section{Preliminaries and auxiliary results}
%%%%%%%%%%%%%%%%%%%%%%%%%%%%%%%%%%%%%%%%%%%%%%%%%%%%%%%%%%%%%%%%%%%%%%

Throughout, $\Xp=\{X_n\}$ denotes a memoryless source
with distribution $P$ that has full support on a finite
alphabet $A$. For $\alpha\in(0,1)$, the tilted distribution
$P_\alpha$ is given by
\bqq
P_\alpha(x) = \frac{ P(x)^\alpha}{Z_\alpha},\quad
x \in A,
\label{eq:Palpha}
\eqq
where the normalising constant 
$Z_\alpha =\sum_{x \in A} P(x)^\alpha$. 
The standard normal cumulative distribution
function is denoted by $\Phi$,
`$\log$' denotes the logarithm taken to base~2, 
and the 
relative entropy between two p.m.f.s $P,Q$ on the same
finite alphabet $A$ is
$D(P \| Q) = \sum_{x \in A} P(x) \log [P(x)/Q(x)].$

\subsection{Normal approximation, compression and hypothesis testing}
%%%%%%%%%%%%%%%%%%%%%%%%%%%%%%%%%%%%%%%%%%%%%%%%%%%%%%%%%%%%%%%%%%%%%%

Here we collect some classical normal approximation
inequalities, a simple one-shot converse bound for data compression
and some useful nonasymptotic results for binary hypothesis testing,
all of which will be employed in the proofs of our results
in Sections~\ref{s:main} and~\ref{s:universal}.

First, we recall the classical 
Berry-Ess\'een bound~\cite{korolev:10,petrov-book:95}
for sums of i.i.d.\ random variables.

\begin{theorem}[Berry-Ess\'een bound]
\label{thm BE bound}
    Let $Z_1, Z_2, \ldots, Z_n$ be i.i.d.\ random variables with mean $\mu= \mathbb{E}(Z_1)$, variance $\sigma^2= \textup{Var}(Z_1)$ and $\rho = \mathbb{E}(| Z_1-\mu |^3) < \infty$. 
Then:
    \begin{equation*}
        \Bigg|   \mathbb{P}\Bigg(  \frac{\sqrt{n}}{\sigma} \Big(\frac{1}{n} \sum_{i=1}^n Z_i - \mu  \Big) \leq x \Bigg) - \Phi(x) \Bigg|  \leq \frac{\rho}{2\sigma^3 \sqrt{n}}, \hspace{0.2in} x \in \mathbb{R}, \hspace{0.07in} n \geq 1.
    \end{equation*}
\end{theorem}

\noindent
The following is a simple corollary of 
Theorem~\ref{thm BE bound}, cf.~\cite{PPV:10,lungu-K:ISIT24,lungu-arxiv:24}.

\begin{lemma}
\label{lem BE exponential}
Let $Z_1,Z_2, \ldots, Z_n$ be as in Theorem~\ref{thm BE bound}. Then:
\begin{equation*}
    \mathbb{E}\Bigg\{  \exp\Big(  -\sum_{i=1}^n Z_i \Big) \mathbb{I}_{ \{ \sum_{i=1}^n Z_i \geq x \}  }     \Bigg\}   \leq \Big(\frac{1}{\sqrt{2\pi}} + \frac{\rho}{\sigma^2} \Big) \frac{1}{\sigma\sqrt{n}} e^{-x}        , \hspace{0.2in} x \in \mathbb{R}, \hspace{0.07in} n \geq 1.
\end{equation*}
\end{lemma}

\noindent
The following one-shot converse bound for
data compression
was established in~\cite{kontoyiannis-verdu:14}.

\begin{theorem}[One-shot converse]
\label{one shot conv thm}
Let $f^*$ be the optimal compressor for 
a finite-valued random variable $X$. Then, for any $k>0$ and $\tau >0$,
    \begin{equation*}
        \mathbb{P}[\ell(f^*(X)) \geq k-1] >  \mathbb{P}[-\log P(X) \geq k + \tau] - 2^{-\tau} .
    \end{equation*}
\end{theorem}
    
\noindent
The next two results were originally developed in the context
of binary hypothesis testing.
Lemma~\ref{lem two distr}~\cite{csiszar:71} can be viewed
as a ``change of measure'' result for hypothesis testing,
and 
Theorem~\ref{thm stein reg conv}~\cite{lungu-K:ISIT24,lungu-arxiv:24,PPV:10}
is a strong, finite-$n$ converse bound for hypothesis testing.

\begin{lemma}[Change of measure]
\label{lem two distr}
Let $P,Q$ be two p.m.f.s with full support on $A$, and let
$S_n$ be an arbitrary subset of $A^n$, 
for some $n\geq 1$.
If
$$\frac{P^n(x^n)}{Q^n(x^n)} \geq \max_{y^n \in S_n}  
\frac{P^n(y^n)}{Q^n(y^n)} \quad\mbox{for all}\;x^n \not \in S_n,$$
then
\begin{equation*}
    P^n(S_n) = \min_{S \subset A^n: Q^n(S^c) \leq Q^n(S_n^c)} P^n(S) .
\end{equation*}
\end{lemma}

\begin{theorem}[Hypothesis testing converse]
    \label{thm stein reg conv}
    Let $P\neq Q$ be p.m.f.s with full support on $A$,
and for $n\geq 1$, $\epsilon\in(0,1)$, let
$$e_1^*(\epsilon) = \min_{S \subset A^n: Q^n(S^c) \leq \epsilon} 
P^n(S).$$ Then, for any $\Delta>0$ such that 
$\epsilon + \frac{B+\Delta}{\sqrt{n}} <1$,
    \begin{equation*}
        \log e_1^*(\epsilon) \geq -n D(Q\|  P) - \sigma \sqrt{n} \Phi^{-1} \Big(\epsilon + \frac{B+\Delta}{\sqrt{n}}\Big) -\frac{1}{2} \log n + \log \Delta,
    \end{equation*}
   where $\sigma^2 = \textup{Var}_Q (\log \frac{Q(X)}{P(X)})$, $\rho = \mathbb{E}_Q|  \log \frac{Q(X)}{P(X)} - \mathbb{E}_Q(\log \frac{Q(X)}{P(X)})|^3$ and $B= \frac{\rho}{2\sigma^3}$.
\end{theorem}

\subsection{Entropy, moments and derivatives identities}
%%%%%%%%%%%%%%%%%%%%%%%%%%%%%%%%%%%%%%%%%%%%%%%%%%%%%%%%%%%%%%%%%

We close with some simple technical lemmas, stated without proof.
Let $X\sim P$ where $P$ has full support on $A$,
and recall the definition of $P_\alpha$ in~(\ref{eq:Palpha}).
The proofs of Lemmas~\ref{lem var, 3rd mom}--\ref{lem entropy deriv}
follow by direct computation; see
for example,~\cite{csiszar:71}
and~\cite{lungu-K:ISIT24,lungu-arxiv:24}.

\begin{lemma}
\label{lem var, 3rd mom}
Define the tilted second and third moments as:
\begin{align*}
\sigma_{2,\alpha}^2
&= 
	\textup{Var}_{P_\alpha} \Big(\log_e \frac{P_\alpha(X)}{P(X)} \Big),\\
\rho_{2,\alpha}
&= 
	\mathbb{E}_{P_\alpha} \Big|  \log_e \frac{P_\alpha(X)}{P(X)} 
	- \mathbb{E}_{P_\alpha} \Big( \log_e \frac{P_\alpha(X)}{P(X)}  
	\Big) \Big|^3 \,,\\
\sigma_{3,\alpha}^2
	&= \textup{Var}_{P_\alpha}(\log_e P(X)),\\
\rho_{3,\alpha}
	&= \mathbb{E}_{P_\alpha} |  \log_e P(X) 
	-  \mathbb{E}_{P_\alpha}(\log_e P(X)) |^3.
\end{align*}
Then, for all $\alpha\in(0,1)$:
$\sigma_{2,\alpha}^2 = (1-\alpha)^2 \sigma_{3,\alpha}^2$
and $\rho_{2,\alpha} = (1-\alpha)^3 \rho_{3,\alpha}$.
\end{lemma}

\begin{lemma}
    \label{lem derivatives}
	For all $\alpha\in(0,1)$ we have the following expressions
	for the derivatives of $D(P_\alpha\|P)$ and $\sigma^2_{3,\alpha}$: 
\begin{align*}
\frac{d D(P_\alpha \| P)}{d \alpha} 
&= 
	(\alpha-1) \sigma_{3,\alpha}^2(\log e),\\
\frac{d^2 D(P_\alpha \| P)}{d \alpha ^2} 
&= 
	(\log e)\sigma_{3,\alpha}^2  + (\log e)(\alpha-1) 
	\frac{d \sigma_{3,\alpha}^2}{d \alpha},\\
\frac{d \sigma_{3,\alpha}^2}{d \alpha} 
&=
	\mathbb{E}_{P_\alpha} \Big[ 
         \Big(\log_e P(X) -  \mathbb{E}_{P_\alpha}[\log_e P(X)] \Big)^3 \Big].
\end{align*}
\end{lemma}

\begin{lemma}
\label{lem D decreasing}
$D(P_\alpha\|  P)$ is a continuous, strictly decreasing function of $\alpha$, 
for $\alpha \in (0,1)$.
\end{lemma}

\begin{lemma}
\label{lem entropy deriv}
	For all $\alpha\in(0,1)$ we have the following expressions
	for the derivatives of $H(P_\alpha)$:
\begin{align*}
\frac{d H(P_\alpha)}{d \alpha} 
&= 
	- (\log e)\alpha \sigma_{3,\alpha}^2,\\
\frac{d^2 H(P_\alpha)}{d \alpha ^2} 
&= 
	-(\log e)\Big[
	\sigma_{3,\alpha}^2 
	+ \alpha \frac{d \sigma_{3,\alpha}^2}{d \alpha} 
	\Big].
\end{align*}
\end{lemma}

\section{Compression of a known source}
%%%%%%%%%%%%%%%%%%%%%%%%%%%%%%%%%%%%%%%%%%%%%%%%%%%%%%%%%%%%%%%%%%%%%%
\label{s:main}

We will now proceed with the first set of 
our main results of this paper, namely 
the development of non-asymptotic achievability and converse bounds 
on the optimal compression rate, $R_n^*(2^{-n\delta},P)$, 
when the excess-rate probability
decays exponentially fast. 

\subsection{Achievability}
%%%%%%%%%%%%%%%%%%%%%%%%%%%%%%%%%%%%%%%%%%%%%%%%%%%%%%%%%%%%%%%%%%%%%%
\label{subsec: achievability}

\begin{theorem}[Achievability]
\label{thm_achievability}
Consider a memoryless source $\Xp$ with p.m.f.\
$P$ with full support on $A$.
Let $\delta \in (0, D(U\|  P))$, where $U$ is the uniform 
distribution over $A$. Then,
for any $n \geq 1$,
    \begin{equation}
        R_n^*(2^{-n\delta },P)  \leq  H(P_{\alpha^*})  - \frac{1}{2(1-\alpha^*)}\frac{\log n}{n} + \frac{c}{n},
\label{eq:achieve}
    \end{equation}
where $P_\alpha$ is defined in~{\em (\ref{eq:Palpha})},
$\alpha^*$ is the unique $\alpha \in (0,1)$ for 
which $\delta=D(P_{\alpha^*}\|  P)$, and
    \begin{equation*}
        c= \log \Bigg(  \frac{1}{\sigma_1} \Big(\frac{1}{\sqrt{2\pi}} + \frac{\rho_1}{\sigma_1^2}\Big)  \Bigg) +  \frac{\alpha^* }{1-\alpha^*} \log \Bigg(  \frac{1}{\sigma_2} \Big(\frac{1}{\sqrt{2\pi}} + \frac{\rho_2}{\sigma_2^2}\Big)    \Bigg),
    \end{equation*}
where $\sigma_1^2= \textup{Var}_{P_{\alpha^*}}(\log_e P_{\alpha^*}(X))$, $\rho_1= \mathbb{E}_{P_{\alpha^*}} |  \log_e P_{\alpha^*}(X) -  \mathbb{E}_{P_{\alpha^*}}(\log_e P_{\alpha^*}(X)) |^3$,
    $\sigma_2^2= \sigma_{2,\alpha^*}^2$ and $\rho_2= \rho_{2,\alpha^*}$, with $\sigma_{2,\alpha}^2$ and $\rho_{2,\alpha}$ defined as in
Lemma~\ref{lem var, 3rd mom}.
\end{theorem}

\noindent
{\bf Remark. }
As noted in the introduction,
the exact bound~(\ref{eq:achieve}) remains
valid in the case of codes with prefix constraints  -- 
i.e., for $R_n^{\sf p}$ in place of $R_n^*$ -- 
with $c+1$ in place of $c$.

\medskip

\noindent
{\sc Proof. }
The proof is based on an explicit construction
of a compressor $f_n$. To that end,
let $0<\beta_n <1$ be a constant that will be chosen later,
and define the set
\begin{equation}
    E_n = 
	\{   x^n \in A^n : \log P^n(x^n) \geq \log \beta_n \}. 
\label{def_E_n}
\end{equation}
Write
$$M_X(\beta_n):=|E_n|,$$ 
for the cardinality of $E_n$,
and suppose that
\begin{equation}
P^n(E_n^c) \leq 2^{-n\delta}.
\label{eq:betatarget}
\end{equation}
To obtain the desired compressor on $A^n$,
first we order the strings $x^n\in E_n$ in order 
of decreasing probability $P^n(x^n)$, breaking
ties arbitrarily, and define $f_n$ via:
\begin{equation*}
  \ell(f_n(x^n)) =
    \begin{cases}
      \lfloor \log k \rfloor & \text{if $x^n$ is the $k^{th}$ element of $E_n$}, \\
      \lceil \log |A|^n \rceil & \text{if $x^n \not\in E_n$}.
    \end{cases}       
\end{equation*}
Note that the longest codeword $f_n(x^n)$ among all strings in $E_n$ 
has length $\lfloor \log M_X(\beta_n) \rfloor$. Therefore, 
for any fixed $\xi\in(0,1)$,
$$
\mathbb{P}[\ell(f_n(X^n)) \geq \lfloor \log M_X(\beta_n) \rfloor +\xi] 
\leq P^n(E_n^c)  \leq 2^{-n\delta}.$$
Hence, by definition,
$nR_n^*(2^{-n\delta },P) \leq \log M_X(\beta_n)+\xi$,
and since $\xi\in(0,1)$ was arbitrary,
\begin{equation*}
    R_n^*(2^{-n\delta },P) \leq \frac{\log M_X(\beta_n)}{n}.
\end{equation*}

The remainder of the proof is devoted to showing that there
is an appropriate $\beta_n$ such that~(\ref{eq:betatarget})
holds, and evaluating the rate $(1/n)\log M_X(\beta_n)$ for 
this choice.

Let $\alpha\in(0,1)$ arbitrary.
We begin by observing that we can express
\begin{align*}
     M_X(\beta_n)
&= \sum_{x^n\in E_n}\frac{P^n_\alpha(x^n)}{P^n_{\alpha}(x^n)}\\
&=  \sum_{x^n \in A^n}  P_\alpha ^n(x^n)  \exp (-\log_e P_\alpha ^n(x^n)) \mathbb{I}_{E_n}(x^n)  \\
%     &=  \sum_{x^n \in A^n}  P_\alpha ^n(x^n)  \exp \Big(-\sum_{i=1}^n \log_e P_\alpha (x_i) \Big) \mathbb{I}\Big \{   \log_e P^n(x^n) \geq \log_e \beta_n \Big\}  \\
    &= \mathbb{ E }_{P_\alpha^n} \Big[   \exp \Big(-\sum_{i=1}^n \log_e P_\alpha (X_i) \Big) \mathbb{I}\Big \{   \sum_{i=1}^n \log_e P (X_i)  \geq \log_e \beta_n \Big\}  \Big].
\end{align*}
We can now apply the version of the 
Berry-Ess\'een bound
in 
Lemma~\ref{lem BE exponential}, 
with $Z_i=\log_e P_\alpha (X_i)$ 
and $x=\alpha \log_e \beta_n - n \log_e Z_\alpha$, to obtain that:
\begin{align}
M_X(\beta_n)
&= 
	\mathbb{ E }_{P_\alpha^n} \Big[   \exp 
	\Big(-\sum_{i=1}^n \log_e P_\alpha (X_i) \Big) 
	\mathbb{I}\Big \{   \sum_{i=1}^n \log_e P_\alpha (X_i)  
	\geq \alpha \log_e \beta_n  - n\log_e Z_\alpha  \Big\}  \Big]
	\nonumber\\
& \leq 
	\frac{1}{\sigma_{1,\alpha}} \Big( \frac{1}{\sqrt{2\pi}} 
	+ \frac{\rho_{1,\alpha}}{\sigma_{1,\alpha}^2} \Big) \frac{1}{\sqrt{n}}
	\exp(-(\alpha \log_e \beta_n  - n\log_e Z_\alpha))    
	\nonumber \\
&=  
	\frac{1}{\sigma_{1,\alpha}} \Big( \frac{1}{\sqrt{2\pi}} 
	+ \frac{\rho_{1,\alpha}}{\sigma_{1,\alpha}^2} \Big) 
	\frac{Z_\alpha^n}{\beta_n^\alpha \sqrt{n}},  
\label{eq_13 upper bound}
\end{align}
where $\sigma_{1,\alpha}^2= \text{Var}_{P_\alpha}(\log_e P_\alpha(X))$ and $\rho_{1,\alpha}= \mathbb{E}_{P_\alpha} |  \log_e P_\alpha(X) -  \mathbb{E}_{P_\alpha}(\log_e P_\alpha(X)) |^3$.

In order to select appropriate values for $\beta_n$ and $\alpha$,
we examine $P^n(E_n^c)$:
\begin{align*}
P^n (E_n^c) 
&= 
	\sum_{x^n \in A^n} P^n(x^n) \mathbb{I}_{E_n^c}(x^n)\\
&=  
	\sum_{x^n \in A^n} P_\alpha^n(x^n) \exp 
	\Bigg(\log_e \frac{P^n(x^n)}{P_\alpha^n(x^n)} \Bigg) 
	\mathbb{I}\{  \log_e   P^n(x^n) < \log_e \beta_n \}\\
&=  
	\sum_{x^n \in A^n} P_\alpha^n(x^n) \exp 
	\Bigg( - \sum_{i=1}^n \log_e \frac{P_\alpha(x_i)}{P(x_i)} \Bigg)
	\mathbb{I}\Bigg\{\sum_{i=1}^n \log_e  P(x_i) < \log_e \beta_n \Bigg\}.
\end{align*}
Rearranging, we can further express
\begin{align*}
P^n (E_n^c) 
&=  
	\sum_{x^n \in A^n} P_\alpha^n(x^n) \exp 
	\Bigg( - \sum_{i=1}^n \log_e \frac{P_\alpha(x_i)}{P(x_i)} \Bigg)\\
&\hspace{1in} 
	\mathbb{I}  \Bigg\{      n\log_e Z_\alpha + \sum_{i=1}^n \log_e  
	P^{1-\alpha}(x_i) < n\log_e Z_\alpha +(1-\alpha) \log_e\beta_n\Bigg\}\\
&=  
	\sum_{x^n \in A^n} P_\alpha^n(x^n) \exp 
	\Bigg( - \sum_{i=1}^n \log_e \frac{P_\alpha(x_i)}{P(x_i)} \Bigg) \\
&\hspace{1in} 
	\mathbb{I}  \Bigg\{      \sum_{i=1}^n \log_e 
	\frac{P_\alpha(x_i)}{P(x_i)} > - n\log_e Z_\alpha -(1-\alpha) 
	\log_e \beta_n \Bigg\},
\end{align*}
so that
\begin{align*}
P^n (E_n^c) 
&=  
	\mathbb{E}_{P_\alpha^n} \Bigg\{ \exp \Bigg( - \sum_{i=1}^n \log_e 
	\frac{P_\alpha(X_i)}{P(X_i)} \Bigg)\\
& \hspace{1in} 
	\mathbb{I} \Bigg\{\sum_{i=1}^n \log_e 
	\frac{P_\alpha(X_i)}{P(X_i)} > - n\log_e Z_\alpha -(1-\alpha) 
	\log_e \beta_n \Bigg\}  \Bigg\}.  
\end{align*}
Applying Lemma~\ref{lem BE exponential} again, 
with $Z_i= \log_e [P_\alpha(X_i)/P(X_i)]$ 
and $x=- n\log_e Z_\alpha -(1-\alpha) \log_e \beta_n$,
yields
\begin{align}
    P^n (E_n^c)  & \leq \frac{1}{\sigma_{2,\alpha}} \Big(  \frac{1}{\sqrt{2\pi}} + \frac{\rho_{2,\alpha}}{\sigma_{2,\alpha}^2} \Big) \frac{1}{\sqrt{n}} \exp \Big( n\log_e Z_\alpha +(1-\alpha) \log_e \beta_n   \Big)    
	\nonumber\\ % \label{berry-ess_exp_form} \\
    &=  \frac{1}{\sigma_{2,\alpha}} \Big( \frac{1}{\sqrt{2\pi}} + \frac{\rho_{2,\alpha}}{\sigma_{2,\alpha}^2}  \Big) \frac{Z_\alpha^n \beta_n^{1-\alpha}}{ \sqrt{n}}.   \label{err_prob}
\end{align} 

Now, setting
$$    \log \beta_n = \frac{1}{1-\alpha} \Bigg( -n \delta + \log \Bigg(  \frac{1}{\frac{1}{\sigma_{2,\alpha}} \Big(\frac{1}{\sqrt{2\pi}} + \frac{\rho_{2,\alpha}}{\sigma_{2,\alpha}^2}  \Big) \frac{Z_\alpha^n }{ \sqrt{n}} }  \Bigg) \Bigg),
$$
immediately yields~(\ref{eq:betatarget}).
Note that, $\beta_n>0$ by~\eqref{err_prob}, 
while~\eqref{def_E_n} and~(\ref{eq:betatarget})
imply $\beta_n<1$. 

As a result, the upper bound for $M_X(\beta_n)$ in~\eqref{eq_13 upper bound}  
becomes:
\begin{align}
    \log M_X(\beta_n) &\leq \log \Bigg(  \frac{1}{\sigma_{1,\alpha}} \Big(\frac{1}{\sqrt{2\pi}} + \frac{\rho_{1,\alpha}}{\sigma_{1,\alpha}^2}\Big) \frac{Z_\alpha^n}{\beta_n^\alpha \sqrt{n}} \Bigg) \nonumber \\
    &= \log \Bigg(  \frac{1}{\sigma_{1,\alpha}} \Big(\frac{1}{\sqrt{2\pi}} + \frac{\rho_{1,\alpha}}{\sigma_{1,\alpha}^2}\Big)  \Bigg)  + n \log Z_\alpha -\frac{1}{2} \log n \nonumber \\
    & \hspace{0.5in} + \frac{\alpha n \delta}{1-\alpha} - \frac{\alpha }{1-\alpha} \log \Bigg(  \frac{1}{\frac{1}{\sigma_{2,\alpha}} \Big(\frac{1}{\sqrt{2\pi}} + \frac{\rho_{2,\alpha}}{\sigma_{2,\alpha}^2} \Big) \frac{Z_\alpha^n }{ \sqrt{n}} }  \Bigg)  \nonumber \\ 
    &= \frac{n}{1-\alpha}(\alpha\delta + \log Z_\alpha )  - \frac{1}{2(1-\alpha)}\log n     \nonumber  \\ 
    & \hspace{0.5in} + \log \Bigg(  \frac{1}{\sigma_{1,\alpha}} \Big( \frac{1}{\sqrt{2\pi}} + \frac{\rho_{1,\alpha}}{\sigma_{1,\alpha}^2} \Big)  \Bigg) + \frac{\alpha }{1-\alpha} \log \Bigg(  \frac{1}{\sigma_{2,\alpha}} \Big( \frac{1}{\sqrt{2\pi}} + \frac{\rho_{2,\alpha}}{\sigma_{2,\alpha}^2} \Big)   \Bigg) . \label{eq_achiev_alpha} 
\end{align}

We finally need to select $\alpha$. 
Since $D(P_\alpha\|  P)$ is continuous and strictly 
decreasing in $\alpha$ (Lemma~\ref{lem D decreasing}),
and since 
$\delta \in (D(P_1\|  P)), D(P_0\|  P)) = (0, D(U\|  P))$,
there exists a unique $\alpha^* \in (0,1)$ 
for which $\delta=D(P_{\alpha^*}\|  P)$. 
For this value of $\alpha$,
\begin{align}
    \alpha^* \delta +\log Z_{\alpha^*}  
&=  \alpha^* \sum_{x \in A} P_{\alpha^*}(x) \log  \frac{P_{\alpha^*}(x) }{P(x)}    + \sum_{x \in A} P_{\alpha^*}(x) \log  \frac{P_{\alpha^*}(x) Z_{\alpha^*}}{P_{\alpha^*}(x)}       \nonumber\\
    &=   \sum_{x \in A} P_{\alpha^*}(x) \log  \frac{P_{\alpha^*}(x)^{\alpha^*} }{P(x)^{\alpha^*}}    + \sum_{x \in A} P_{\alpha^*}(x) \log  \frac{P(x)^{\alpha^*}}{P_{\alpha^*}(x)}   \nonumber\\
    &=  \sum_{x \in A} P_{\alpha^*}(x) \log\Big(   P_{\alpha^*}(x)^{(\alpha^*-1)} \Big)         \nonumber\\
    &= (1- \alpha^*)  H(P_{\alpha^*}).
	\label{eq:PalphastarZ}
\end{align}

Substituting this in~\eqref{eq_achiev_alpha} 
yields~(\ref{eq:achieve}), as required.
\qed

\subsection{Converse}
%%%%%%%%%%%%%%%%%%%%%%%%%%%%%%%%%%%%%%%%%%%%%%%%%%%%%%%%%%%%%%%%%%%%%%
\label{subsec: converse}

\begin{theorem}[Converse]
\label{thm_converse}
Consider a memoryless source $\Xp$ with p.m.f.\ $P$
with full support on $A$. Let $\delta \in (0, D(U\|  P))$, 
where $U$ is the uniform distribution over $A$. 
Then, for any $n > N_0$,
    \begin{equation}
        R_n^*(2^{-n\delta },P)  \geq  H(P_{\alpha^*})  - \frac{1}{2(1-\alpha^*)}\frac{\log n}{n} - \frac{C}{n},
\label{eq:converse}
    \end{equation}
where $P_\alpha$ is as defined in~{\em (\ref{eq:Palpha})},
$\alpha^*$ is the unique $\alpha \in (0,1)$ for 
which $\delta=D(P_{\alpha^*}\|  P)$, and,
    \begin{align*}
        N_0 &= \max \Bigg\{  4.4 \Big( \frac{\hat{\rho}_3}{ \tilde{\sigma}_3^3 }  + 1  \Big)^2, \frac{4(1+q+r)^2}{p^2}, \frac{2(1+q+r)}{p(1- \alpha^*)}, N_1, N_2 \Bigg\}, \\
        C &= \frac{1+q+r}{1-\alpha^*}  + \frac{\log e}{2} \Big| \tilde{\sigma}_3^2 - (1-\alpha^*) \hat{\rho}_3\Big|   + \frac{\log e}{2}  
(\hat{\sigma}_3 ^2 + \hat{\rho}_3) +1,
    \end{align*}
where $\sigma_{3,\alpha}$ and $\rho_{3,\alpha}$ are as defined in 
Lemma~\ref{lem var, 3rd mom}, and the remaining constants are given by:
\begin{align*}
&\sigma_3^2=\sigma_{3,\alpha^*}^2,\;\;
\tilde{\sigma}_3^2 = \inf_{\alpha \in (0,1)} \sigma_{3,\alpha}^2,\;\;
\hat{\sigma}_3^2 = \sup_{\alpha \in (0,1)} \sigma_{3,\alpha}^2,\;\;
\hat{\rho}_3 = \sup_{\alpha \in (0,1)} \rho_{3,\alpha},\\
&p =  (\log e)(1-\alpha^*) \sigma_3 ^2,\;\;  
q= \frac{\log e}{2}(\hat{\sigma}_3 ^2+ \hat{\rho}_3),\;\;
r= 19 (\log e) (1-\alpha^*)  \Big( \frac{\hat{\rho}_3}{ \tilde{\sigma}_3^3 }  + 1  \Big)  \sqrt{\sigma_3 ^2 + \hat{\rho}_3}, \\
&N_1=\min\big\{n\geq 8\;:\;\log n\leq p\sqrt{n}\big\},
\;\;
N_2=\min\big\{n\geq 3\;:\;\log n\leq p(1-\alpha^*)n\big\}.
\end{align*}
\end{theorem}

\noindent
{\bf Remark. }
Again, in view of the remark at the end of
Section~\ref{s:VRDC},
the bound~(\ref{eq:converse}) remains
valid for prefix free codes
exactly as stated.

\medskip

\noindent
{\sc Proof. }
Taking $k=nR$, for some $R>0$, letting $\tau = n \delta$, 
and replacing $X$ with $X^n$ (and $f^*$ with $f_n^*$) in
the one-shot converse in Theorem~\ref{one shot conv thm},
yields:
\begin{equation*}
    \mathbb{P}\Big[\ell(f_n^*(X^n))\geq n\Big(R-\frac{1}{n}\Big) \Big] 
	> \mathbb{P}[-\log P^n(X^n) \geq n(R + \delta)] - 2^{-n\delta} .
\end{equation*}
Therefore, if we can find a rate $R$ for which
\begin{equation*}
    \mathbb{P}[\log P^n(X^n) \leq -n(R + \delta)]  \geq 2 \times 2^{-n\delta},  % \label{conv_goal}
\end{equation*} 
this would imply that 
$\mathbb{P}[\ell(f_n^*(X^n)) \geq n(R-1/n)] > 2^{-n\delta}$,
which would in turn give
\bqq
R_n^*(2^{-n\delta},P)\geq R-\frac{1}{n}.
\label{eq:firstRbound}
\eqq
Writing
\begin{equation*}
    S_n = \{  x^n \in A^n : \log P^n(x^n) \leq -n(R + \delta)\},
\end{equation*}
our objective then is to find an appropriate rate $R$ 
for which
\begin{equation}
    P^n(S_n) \geq 2 \times 2^{-n\delta}. \label{eq:conv_goal}
\end{equation}
Much of the remainder of the proof will be devoted to
establishing~(\ref{eq:conv_goal}).

First we will use Lemma~\ref{lem two distr} to express 
$P^n(S_n)$ in terms of $P_{\alpha_n}^n(S_n^c)$ 
for some $\alpha_n \in (0,1)$ that will be chosen later. 
Indeed, since
\begin{align}
    S_n % &= \Bigg\{  x^n \in A^n : \sum_{i=1}^n \log P(x_i) \leq -n(R + \delta)  \Bigg\}  \nonumber \\
    &= \Bigg\{  x^n \in A^n : \sum_{i=1}^n \log P^{1-\alpha_n}(x_i) \leq -(1-\alpha_n)n(R + \delta)  \Bigg\}  \nonumber \\
    &= \Bigg\{  x^n \in A^n : \log \frac{ P^n(x^n)}{P^n_{\alpha_n}
	(x^n)} \leq -(1-\alpha_n)n(R + \delta) + n \log Z_{\alpha_n} \Bigg\},  \label{eq_S_n_conv}
\end{align}
we have that
$$\frac{P^n(x^n)}{P_{\alpha_n}^n(x^n)} \geq \max_{x^n \in S_n}   
\frac{P^n(x^n)}{P_{\alpha_n}^n(x^n)}\quad\mbox{for all}\; 
x^n \not\in S_n,$$
so we can apply Lemma~\ref{lem two distr} to obtain
\begin{equation}
	P^n(S_n) = \min_{S \subset A^n: P_{\alpha_n}^n(S^c) \leq P_{\alpha_n}^n(S_n^c) } P^n(S) .
    \label{P and P_alpha_n conncetion}
\end{equation}

The next step will be an application of the Berry-Ess\'een bound
to obtain an upper bound on
$P_{\alpha_n}^n(S_n^c)$. From~\eqref{eq_S_n_conv} we get
\begin{align}
    P_{\alpha_n}^n(S_n^c) 
    = P_{\alpha_n}^n\Bigg(\sum_{i=1}^n \log_e \frac{ P_{\alpha_n}(x_i)}{P(x_i)} < \frac{ (1-\alpha_n)n(R + \delta) -  n \log Z_{\alpha_n} }{\log e}   \Bigg). 
\label{conv_BE_1}   
\end{align}
Choosing $R$ as
\begin{equation}
     R = \frac{1}{1-\alpha_n} \Big[ \mathbb{E}_{P_{\alpha_n}} 
\Big(  \log \frac{ P_{\alpha_n}(X)}{P(X)}\Big) + \log Z_{\alpha_n} \Big] 
-\delta,
\label{eq_find_R}
\end{equation}
we have
\begin{equation*}
    \frac{ (1-\alpha_n)n(R + \delta) -  n \log Z_{\alpha_n} }{\log e} = n \mathbb{E}_{P_{\alpha_n}} \Big(  \log_e \frac{ P_{\alpha_n}(X)}{P(X)}\Big),
\end{equation*}
and the Berry-Ess\'een in Theorem~\ref{thm BE bound} 
applied to~(\ref{conv_BE_1}) yields
\begin{align*}
    P_{\alpha_n}^n(S_n^c) = P_{\alpha_n}^n\Bigg(\sum_{i=1}^n \log_e \frac{ P_{\alpha_n}(x_i)}{P(x_i)} < n \mathbb{E}_{P_{\alpha_n}} \Big(  \log_e \frac{ P_{\alpha_n}(X)}{P(X)}\Big) \Bigg) 
    \leq \Phi(0) + \frac{\rho_{2,\alpha_n}}{2 \sigma_{2,\alpha_n}^3 \sqrt{n}}.
\end{align*}
Using Lemma~\ref{lem var, 3rd mom}, the bound becomes
\begin{align}
P_{\alpha_n}^n(S_n^c) \leq \frac{1}{2} + \frac{ \frac{\rho_{2,\alpha_n}}{(1-\alpha_n)^3} }{  \frac{2 \sigma_{2,\alpha_n}^3 \sqrt{n}}{(1-\alpha_n)^3} }  = \frac{1}{2} + \frac{\rho_{3,\alpha_n}}{2 \sigma_{3,\alpha_n}^3 \sqrt{n}} \leq   \frac{1}{2} + \frac{\hat{\rho}_3}{2 \tilde{\sigma}_3^3 \sqrt{n}},  \label{rho sigma algebra}
\end{align}
where $\sigma_{3,\alpha_n}$ and $\rho_{3,\alpha_n}$ are as in 
Lemma~\ref{lem var, 3rd mom},
 $\tilde{\sigma}_3^2 := \inf_{\alpha \in (0,1)} \sigma_{3,\alpha}^2 $    and 
      $\hat{\rho}_3 := \sup_{\alpha \in (0,1)} \rho_{3,\alpha} $.

Using this bound, from~\eqref{P and P_alpha_n conncetion} we get
\begin{equation*}
    P^n(S_n) \geq \min_{S \subset A^n: P_{\alpha_n}^n(S^c) \leq   \frac{1}{2} + \frac{\hat{\rho}_3}{2 \tilde{\sigma}_3^3 \sqrt{n}}} P^n(S) .
\end{equation*}
Then, using the hypothesis testing converse in 
Theorem~\ref{thm stein reg conv}, with $P_{\alpha_n}$ in the place of $Q$, 
$\epsilon = \frac{1}{2} + \frac{\hat{\rho}_3}{2 \tilde{\sigma}_3^3 \sqrt{n}}$,
and $\Delta=1$, yields
\begin{align}
    \log P^n(S_n) & \geq -n D(P_{\alpha_n} \| P) -  \sigma_{2,\alpha_n} (\log e) \sqrt{n} \Phi^{-1} \Big(\frac{1}{2} 
+\frac{\hat{\rho}_3}{2 \tilde{\sigma}_3^3 \sqrt{n}} + \frac{\rho_{2,\alpha_n}}{2 \sigma_{2,\alpha_n}^3 \sqrt{n}} + \frac{1}{\sqrt{n}} \Big) -\frac{1}{2} \log n \label{stein reg applied}\\
    & \geq -n D(P_{\alpha_n} \| P) - \sigma_{2,\alpha_n} (\log e) \sqrt{n}  \Phi^{-1} \Big(\frac{1}{2} + \frac{\hat{\rho}_3}{ \tilde{\sigma}_3^3 \sqrt{n}}  + \frac{1}{\sqrt{n}} \Big)  - \frac{1}{2} \log n  \label{invcdf_increasing},
\end{align}
as long as $ \frac{1}{2} +\frac{\hat{\rho}_3}{2 \tilde{\sigma}_3^3 \sqrt{n}} 
+ \frac{\rho_{2,\alpha_n}}{2 \sigma_{2,\alpha_n}^3 \sqrt{n}} 
+ \frac{1}{\sqrt{n}} <1 $, i.e., for $n > 4(\frac{\hat{\rho}_3}{2 
\tilde{\sigma}_3^3 } + \frac{\rho_{2,\alpha_n}}{2 \sigma_{2,\alpha_n}^3} + 1)^2 $. Note that $\frac{\hat{\rho}_3}{2 \tilde{\sigma}_3^3 } \geq \frac{\rho_{2,\alpha_n}}{2 \sigma_{2,\alpha_n}^3}$ by~\eqref{rho sigma algebra}, which 
justifies~\eqref{invcdf_increasing} and also 
means that~\eqref{stein reg applied}  still holds when $n> 
4(\frac{\hat{\rho}_3}{\tilde{\sigma}_3^3 } + 1)^2 $.
   
%%%%%%%%%%%%%%%%%%%%%%%%%%%%%%%%%%%%%%%%%%%%%%%%%%

Recall that the existence of a unique $\alpha^*$ as in the statement 
of the theorem is already established in Theorem~\ref{thm_achievability}.
In order to bound $\log P^n(S_n)$ below further,
we will use simple Taylor expansions to get bounds 
on $D(P_{\alpha_n}\|P)$, $\sigma_{2,\alpha_n}$
and the $\Phi^{-1}$ term in~(\ref{invcdf_increasing}).
First we expand $D(P_{\alpha_n}\|P)$ around $\alpha^*$.
Using the expressions for the derivatives 
of $D(P_\alpha\|P)$ in Lemma~\ref{lem derivatives}, 
gives
\begin{align}
D(P_{\alpha_n} \| P) 
&= 
	D(P_{\alpha^*} \| P) 
	+ (\alpha_n - \alpha^*) (\alpha^* -1)\sigma_3^2(\log e)
	\nonumber\\
&\quad + \frac{\log e}{2} (\alpha_n - \alpha^*)^2 
   \Bigg( \sigma_{3,u}^2  +  (u-1) \mathbb{E}_{P_u} \Big[ 
         \Big(\log_e P(X) -  \mathbb{E}_{P_u}(\log_e P(X)) \Big)^3 \Big]    \Bigg)  \label{eq taylor D(P alpha, P)},
\end{align}
for some $u \in [\alpha^*, \alpha_n]$, 
assuming $\alpha^* \leq  \alpha_n $, 
and where $\sigma_3= \sigma_{3,\alpha^*}$.
Letting $\hat{\sigma}_3 ^2 = \sup_{\alpha \in (0,1)} \sigma_{3,\alpha}^2$, 
we have $\sigma_{3,u}^2 \leq \hat{\sigma}_3 ^2$,
and we observe that $|u-1| \leq 1$ and that
\begin{align*}
     \Big| \mathbb{E}_{P_u} \Big[ 
         \Big(\log_e P(X) -  \mathbb{E}_{P_u}(\log_e P(X)) \Big)^3 \Big] \Big| \leq \mathbb{E}_{P_u} \Big| \log_e P(X) -  \mathbb{E}_{P_u}(\log_e P(X)) \Big|^3 \leq \hat{\rho}_3.
\end{align*}
Therefore,
\begin{align}
 D(P_{\alpha_n} \| P) \leq D(P_{\alpha^*} \| P) + (\alpha_n - \alpha^*) (\alpha^* -1) \sigma_3 ^2 (\log e) + \frac{\log e}{2} (\alpha_n - \alpha^*)^2 (\hat{\sigma}_3 ^2+ \hat{\rho}_3).  \label{bound D(P alpha || P)}
\end{align}
Next, since $\sigma_{2,\alpha_n} = (1-\alpha_n) \sigma_{3,\alpha_n}$ 
by Lemma~\ref{lem var, 3rd mom}, 
we can expand $\sigma_{3,\alpha_n}$ instead of $\sigma_{2,\alpha_n}$
around $\alpha^*$, 
yielding
\begin{align*}
   \sigma_{3,\alpha_n}^2 = \sigma_3 ^2 + (\alpha_n - \alpha^*) \mathbb{E}_{P_w} \Big[ 
         \Big(\log_e P(X) -  \mathbb{E}_{P_w}(\log_e P(X)) \Big)^3 \Big] \leq \sigma_3 ^2 + \hat{\rho}_3 ,
\end{align*}
for some $w \in [\alpha^*, \alpha_n]$, 
assuming $\alpha_n - \alpha^* \leq 1 $.
Hence,
\begin{equation}
    \sigma_{2,\alpha_n} = (1-\alpha_n) \sigma_{3,\alpha_n} \leq  (1-\alpha^*) \sqrt{\sigma_3 ^2 + \hat{\rho}_3} .
  \label{bound sigma}
\end{equation}
And for the third term, we expand $\Phi^{-1}$ around $1/2$
to obtain that
\begin{align}
    \Phi^{-1} \Big(\frac{1}{2} + \frac{\hat{\rho}_3}{ \tilde{\sigma}_3^3 \sqrt{n}}  + \frac{1}{\sqrt{n}} \Big)  &= \Phi^{-1} \Big(\frac{1}{2}\Big) + \Big( \frac{\hat{\rho}_3}{ \tilde{\sigma}_3^3 \sqrt{n}}  + \frac{1}{\sqrt{n}}  \Big) \frac{1}{\phi(\Phi^{-1}(x))} \nonumber \\
    & \leq  \frac{1}{\sqrt{n}} \Big( \frac{\hat{\rho}_3}{ \tilde{\sigma}_3^3 }  + 1  \Big) \frac{1}{\phi(\Phi^{-1}(\frac{1}{2} + \frac{\hat{\rho}_3}
{ \tilde{\sigma}_3^3 \sqrt{n}}  + \frac{1}{\sqrt{n}}))} \nonumber \\
    & \leq  \frac{1}{\sqrt{n}} \Big( \frac{\hat{\rho}_3}{ \tilde{\sigma}_3^3 }  + 1  \Big) \frac{1}{\phi(2)} \leq  \frac{19}{\sqrt{n}} \Big( \frac{\hat{\rho}_3}{ \tilde{\sigma}_3^3 }  + 1  \Big) , \label{phi bound}
\end{align}
for some 
$x \in (\frac{1}{2}, \frac{1}{2} + \frac{\hat{\rho}_3}{ \tilde{\sigma}_3^3 
\sqrt{n}}  + \frac{1}{\sqrt{n}} )$,
where~\eqref{phi bound} holds as long as:
$$n \geq 
4.4  \Big( \frac{\hat{\rho}_3}{ \tilde{\sigma}_3^3 }  + 1  \Big)^2 >
\frac{( \frac{\hat{\rho}_3}{ \tilde{\sigma}_3^3 }  + 1  )^2}{(\Phi(2)-1/2)^2}.$$
Finally, substituting the bounds~\eqref{bound D(P alpha || P)},~\eqref{bound sigma} and~\eqref{phi bound} into~\eqref{invcdf_increasing}, gives
\begin{align}
     \log P^n(S_n)
    & \geq -n D(P_{\alpha^*} \| P) - n (\alpha_n - \alpha^*) (\log e) (\alpha^* -1) \sigma_3 ^2 - \frac{n \log e}{2} (\alpha_n - \alpha^*)^2 (\hat{\sigma}_3 ^2+ \hat{\rho}_3) \nonumber \\
    & \hspace{0.6in} -(1-\alpha^*) \sqrt{\sigma_3 ^2 + \hat{\rho}_3} \times (\log e) \sqrt{n} \times \frac{19}{\sqrt{n}} \Big( \frac{\hat{\rho}_3}
{ \tilde{\sigma}_3^3 }  + 1  \Big) - \frac{1}{2} \log n
	\nonumber \\ 
    & \geq -n \delta + p n (\alpha_n - \alpha^*) - q n (\alpha_n - \alpha^*)^2 -r - \frac{1}{2} \log n, 
	\label{eq:prechoice}
\end{align}
where $\delta,p,q$ and $r$ are as in the statement of the theorem.

Now we are in a position to choose $\alpha_n$. Letting
\begin{equation}
    \alpha_n = \alpha^* + \frac{1}{2p}  \frac{\log n}{n} 
	+ \frac{1+q+r}{p}  \frac{1}{n},  
\label{alpha n}
\end{equation}
simple algebra shows that we have
\begin{equation*}
     p n (\alpha_n - \alpha^*) - q n (\alpha_n - \alpha^*)^2 -r 
- \frac{1}{2} \log n \geq 1, 
% \label{ineq for alpha n}
\end{equation*}
as long as $n$ is large enough so that
\bqq
\frac{\log n}{\sqrt{n}} \leq p \quad\mbox{and}\quad
n \geq \frac{4(1+q+r)^2}{p^2},
\label{eq:nconditions}
\eqq
in which case~(\ref{eq:prechoice}) becomes
$\log P^n(S_n)\geq -n\delta +1$, which is exactly
our desired bound~\eqref{eq:conv_goal}.
To ensure its validity, we must also ensure that 
the earlier conditions we applied -- namely, that
$\alpha_n \geq \alpha^*$, that $\alpha_n - \alpha^* \leq 1$ and 
that $\alpha_n \in (0,1)$ -- are also satisfied. 
The inequalities $\alpha_n >0 $ and 
$\alpha_n \geq \alpha^*$ follow from~\eqref{alpha n},
$\alpha_n - \alpha^* <1$ follows 
from~\eqref{alpha n}
and~(\ref{eq:nconditions}), and finally, 
if we further take $n$ large enough that
$$\frac{\log n}{n}< p(1-\alpha^*)\quad\mbox{and}\quad
n > \frac{2(1+q+r)}{p(1- \alpha^*)},$$
then $\alpha_n < 1$ easily follows from~\eqref{alpha n}.
To summarize, taking $n>N_0$ as in the statement of the theorem
ensures the validity of our key bound~(\ref{eq:conv_goal}).

The final step of the proof consists of showing that,
with $R$ chosen as in~(\ref{eq_find_R})
and with $\alpha_n$ as in~(\ref{alpha n}),
the rate $R$ can be bounded below 
so that, in combination with~(\ref{eq:firstRbound}),
we will obtain the desired bound~(\ref{eq:converse})
in the statement of the theorem.

To that end, starting from the definition of $R$ in~\eqref{eq_find_R}, 
we can re-write
\begin{align}
R  &= \mathbb{E}_{P_{\alpha_n}} \Big(  \log \frac{ P_{\alpha_n}(X)}{P(X)}\Big) + \frac{1}{1-\alpha_n} \Big[  \alpha_n \mathbb{E}_{P_{\alpha_n}} \Big(  \log \frac{ P_{\alpha_n}(X)}{P(X)}\Big) + \log Z_{\alpha_n} \Big] -\delta \nonumber \\
    &= D(P_{\alpha_n}\|  P) + \frac{1}{1-\alpha_n} \Big[  \alpha_n D(P_{\alpha_n}\|  P) +(1-\alpha_n)H(P_{\alpha_n})- \alpha_n D(P_{\alpha_n}\|  P) \Big] -\delta \nonumber \\
    &= D(P_{\alpha_n}\|  P)+ H(P_{\alpha_n}) -\delta .  
\label{expression R before taylor}
\end{align}
To bound this below, we will expand
$D(P_{\alpha_n}\|P)$ and
$H(P_{\alpha_n})$ around $\alpha^*$. 
For $D(P_{\alpha_n}\|P)$, continuing from 
the Taylor expansion in~\eqref{eq taylor D(P alpha, P)} we have
\begin{align}
   D(P_{\alpha_n}\|  P) 
         & \geq \delta - (\alpha_n - \alpha^*) (1- \alpha^* ) \sigma_3 ^2 (\log e) - \frac{\log e}{2} (\alpha_n - \alpha^*)^2 \Big| \tilde{\sigma}_3^2 -(1-\alpha^*)\hat{\rho}_3\Big| \label{eq taylor D 2}  \\
         & \geq \delta - (\alpha_n - \alpha^*)(1- \alpha^* ) \sigma_3 ^2 (\log e) - \frac{\log e}{2n} \Big| \tilde{\sigma}_3^2 -(1-\alpha^*)\hat{\rho}_3\Big| \label{eq taylor D 3} \\
         & = \delta - \frac{\log n}{2n} - \frac{1+q+r}{n} - \frac{\log e}{2n} \Big| \tilde{\sigma}_3^2 -(1-\alpha^*)\hat{\rho}_3\Big| ,  \label{eq taylor D 4}
\end{align}
where~\eqref{eq taylor D 2} holds because the right-hand side
is a lower bound to~\eqref{eq taylor D(P alpha, P)} since
$\sigma_{3,u}^2 \geq \tilde{\sigma}_3^2$, $1-u \leq 1-\alpha^*$ and 
by the definition of $\hat{\rho}_3$;~\eqref{eq taylor D 3} follows from 
the definition of $\alpha_n$ and the fact that $n$ is taken sufficiently large as above;
and~\eqref{eq taylor D 4} follows by substituting the expression of 
$\alpha_n$ in~\eqref{alpha n}.
Similarly expanding $H(P_{\alpha_n})$ 
around $\alpha^*$ and using the expressions for the derivatives 
of $H(P_\alpha)$ in~Lemma \ref{lem entropy deriv}, 
\begin{align}
    H(P_{\alpha_n})&= H(P_{\alpha^*})  
	- (\alpha_n - \alpha^*) \alpha^* \sigma_3 ^2  \log e 
	\nonumber\\
    & \hspace{0.5in} 
	- \frac{\log e}{2} (\alpha_n - \alpha^*)^2 
	\Bigg( \sigma_{3,v}^2  +  v  \mathbb{E}_{P_v} \Big[ 
     \Big(\log_e P(X) -  \mathbb{E}_{P_v}(\log_e P(X)) \Big)^3 \Big]    \Bigg)  \label{eq taylor H 1} \\
    & \geq H(P_{\alpha^*})  - (\alpha_n - \alpha^*) \alpha^* \sigma_3 ^2  \log e  - \frac{\log e}{2} (\alpha_n - \alpha^*)^2 (\hat{\sigma}_3 ^2 +1 \times 
\hat{\rho}_3)  \label{eq taylor H 2} \\
    & \geq H(P_{\alpha^*})  - (\alpha_n - \alpha^*) \alpha^* \sigma_3 ^2  \log e  - \frac{\log e}{2n}  (\hat{\sigma}_3 ^2 + \hat{\rho}_3)  \label{eq taylor H 3} \\
    & = H(P_{\alpha^*})  - \frac{\alpha^*}{2(1-\alpha^*)} \frac{\log n}{n}  - \frac{\alpha^*(1+q+r)}{1-\alpha^*} \frac{1}{n} - \frac{\log e}{2n}  
(\hat{\sigma}_3 ^2 + \hat{\rho}_3)  \label{eq taylor H 4},
\end{align}
where~\eqref{eq taylor H 1} holds for some 
$v \in [\alpha^*, \alpha_n]$;~\eqref{eq taylor H 2} holds 
since $\sigma_{3,v}^2 \leq \hat{\sigma}_3^2$, $v \leq 1$ and 
by the definition of $\hat{\rho}_3$;~\eqref{eq taylor H 3} follows 
from the definition of $\alpha_n$ and the fact that $n$ is taken sufficiently large as above; 
and~\eqref{eq taylor H 4} 
follows by substituting the expression of $\alpha_n$ in~\eqref{alpha n}.

Substituting the bounds~\eqref{eq taylor D 4} 
and~\eqref{eq taylor H 4} into~\eqref{expression R before taylor}, 
gives
\begin{align*}
    R % & \geq H(P_{\alpha^*}) -\frac{1}{2}\Big(1+\frac{\alpha^*}{1-\alpha^*}\Big)\frac{\log n}{n} - \Bigg\{  (1+q+r)\Big(1+\frac{\alpha^*}{1-\alpha^*}\Big) +   \nonumber \\
    % & \hspace{1in} + \frac{\log e}{2} \Big| \tilde{\sigma}_3^2 -  (1-\alpha^*)\hat{\rho}_3\Big|   + \frac{\log e}{2}  (\hat{\sigma}_3 ^2 + \hat{\rho}_3)  \Bigg\} \times \frac{1}{n \\
     \geq H(P_{\alpha^*}) & -\frac{1}{2(1-\alpha^*)} \frac{\log n}{n}\\
& - \Bigg\{  \frac{1+q+r}{1-\alpha^*} 
   + \frac{\log e}{2} \Big| \tilde{\sigma}_3^2 - (1-\alpha^*) \hat{\rho}_3\Big|   + \frac{\log e}{2}  (\hat{\sigma}_3 ^2 + \hat{\rho}_3)  \Bigg\} \times \frac{1}{n} , 
\end{align*}
and combining this with~(\ref{eq:firstRbound}) yields~(\ref{eq:converse})
and completes the proof.
\qed

\section{The number of low-entropy types}
%%%%%%%%%%%%%%%%%%%%%%%%%%%%%%%%%%%%%%%%%%%%%%%%%%%%%%%%%%%%%%%%%%%%%%
\label{s:types}

In this section we prove a fine combinatorial estimate
on the number of stings whose empirical entropy is below
a given threshold. This  will be important in the
proof of the universality result we derive in 
Section~\ref{s:universal}, and it may also be of 
independent interest.

Let $A$ be a fixed finite alphabet of size $|A|=m\geq 2$.
We write $\clP$ for the space of p.m.f.s $P$ on $A$
and denote by $\clP_n\subset\clP$
the subset of all $n$-types on $A$;
recall that $P\in\clP$ is an $n$-type if,
for each $a\in A$,
$P(a)$ is of the form $k/n$ for an integer
$0\leq k\leq n$.
For a string $x^n\in A^n$, we write $\hat{P}_{x^n}$
for the {\em type} of $x^n$, i.e., for the
empirical p.m.f.\ it induces on $A$:
$$\hat{P}_{x^n}(a)=\frac{1}{n}\sum_{i=1}^n\IND_{\{x_i=a\}},
\quad a\in A.$$
Clearly, $\hat{P}_{x^n}$ is always an $n$-type.

Our main result in this section gives an accurate
estimate of the number of strings $x^n$ whose
types $\hat{P}_{x^n}$ have entropy no greater than a given threshold.
Recall that the notation $a_n=\Theta(b_n)$ for two 
nonnegative sequences
$a_n$ and $b_n$ means that there are finite
nonzero constants $c,c'$ such that
$c b_n\leq a_n\leq c' b_n$ for all $n$.

\begin{theorem}[Low-entropy types]
\label{thm:Theta}
Let $Q$ be a non-uniform p.m.f.\ of full support on $A$, and 
for each $n\geq 1$ consider
the set
\bqq
B_n=\big\{x^n\in A^n:H(\hat{P}_{x^n})\leq H(Q)\big\}.
\label{eq:Bn}
\eqq
Then the cardinality $|B_n|$ of $B_n$ satisfies:
\bqq
|B_n|=\Theta\Big(n^{\frac{m-3}{2}}2^{nH(Q)}
\Big),
\label{eq:Theta}
\eqq
where the implied constants depend on $Q,m$.
\end{theorem}

Theorem~\ref{thm:Theta} follows immediately from
the lower bound in Theorem~\ref{thm:LB}
combined with the upper bound in Theorem~\ref{thm:UB},
established in Sections~\ref{s:LB} and~\ref{s:UB},
respectively. These results also provide additional
information on how the implied constants
in~(\ref{eq:Theta}) depend on the problem parameters.
In fact, the upper bound in Theorem~\ref{thm:UB}
is shown to hold uniformly
in a neighbourhood of $Q$.

\subsection{Preliminaries}
%%%%%%%%%%%%%%%%%%%%%%%%%%%%%%%%%%%%%%%%%%%%%%%%%%%%%%%%%%%%%%%%%%%%%%

For an $n$-type $P\in\clP_n$, we write
$\mathcal{T}(P)=\{x^n\in A^n:\hat{P}_{x^n}=P\}$ for the {\em type class} of $P$,
and we denote the support of 
any $P \in \mathcal{P}$ as $\mathrm{supp}(P)$.

The following useful notation will be used throughout
this section.
For two expressions $a$ and $b$ that may depend 
on $n$, $Q$, and possibly on other parameters, we write
$a \gtrsim_Q b$ to indicate that that there exists a constant 
$K = K(Q)$ that depends on $Q$ such that $a \geq K(Q) \cdot b$.
Similarly,
$a \approx b$ means $a \lesssim b$ and $b \lesssim a$. 
When no subscripts are given in $\gtrsim,\lesssim$ and $\approx$,
the implied constants are absolute.
The big-$O$ and small-$o$ notation is used as usual, and again we 
use subscripts to denote the dependence of the implied constants
on the problem parameters. For example, for a positive function 
$g$, $f(n) = O_Q(g(n))$ means that there exists $K = K(Q)$ such that 
$|f(n)| \leq K(Q)g(n)$ for all $n$ large enough. 

The derivations of Theorems~\ref{thm:LB} and~\ref{thm:UB}
involve geometric considerations regarding certain subsets
of $\clP$. We will use the following terminology.
A {\em convex body} is a convex, open, non-empty, and bounded 
subset of $\R^d$. For a convex body $C,$ we denote its volume 
by $\mathrm{Vol}(C)$ and its surface area by $\mathrm{Sur}(C).$ 
The Minkowski sum of two subsets $C_1,C_2$ or $\RL^d$ is
$C_1+C_2=\{x+y:x\in C_1,y\in C_2\}$. Writing $B_2(\epsilon)$
for the closed Euclidean ball of radius $\epsilon>0$ 
centred at the origin,
the Minkowski-Steiner formula~\cite{federer:book}
says that the surface area of a convex body can be
expressed as
$$\mathrm{Sur}(C) =
 \lim_{\epsilon \downarrow0}\frac{\mathrm{Vol}(C+B_2(\epsilon))
- \mathrm{Vol}(C)}{\epsilon}.$$
Moreover,
the volume $\mathrm{Vol}(C+B_2(\epsilon))$ is locally 
a polynomial of degree $d$ in $\epsilon$.

A {\em full-rank lattice} $\Gamma$ in $\RL^d$
is a discrete
additive subgroub $\Gamma$ of $\RL^d$ generated by 
$d$ linearly independent elements
$v_1,\ldots,v_d$ of $\RL^d$.
The lemma below is a standard result on counting lattice 
points in a convex body; see~\cite[Lemma~3.22]{TV06:book}.

\begin{lemma}
\label{taovu}
Let $\Gamma \subset\mathbb{R}^d$ be a full-rank lattice
generated by $\{v_1,\ldots,v_d\}$,
and let $B\subset\mathbb{R}^d$ be a convex body.
Then for $R > 0$,
\[
| (R \cdot B)\cap \Gamma|
= (R^d + O_{\Gamma,B,d}(R^{d-1}))\frac{\operatorname{Vol}(B)}{\mathrm{Vol}(v_1\wedge\ldots\wedge v_d)},
\]
where $v_1\wedge\ldots\wedge v_d$ denotes the parallelepiped 
with vertices $v_1,\ldots,v_d$,
and the 
$O_{\Gamma,B,d}(R^{d-1})$ term is bounded
in absolute value by $K R^{d-1}$, for
some constant $K$ that depends
on $\Gamma,B$ and $d$.
\end{lemma}

The next lemma is a standard result, which can be obtained 
by using the Robbins refinement of the Stirling approximation 
in the multinomial coefficient; see, e.g.,~\cite{csiszar:book2}.

\begin{lemma}[Size of type classes]
\label{lem:stirling}
Let $P$ be an $n$-type with
full support on an alphabet of size $k$.
Then
\[
|\clT(P)|
\approx_k 
\frac{2^{nH(P)}}{n^{(k-1)/2}}\,
\prod_{a\in A} \frac{1}{\sqrt{P(a)}},
\]
where the implied constants depend only on $k$.
\end{lemma}

\subsection{Lower bound}
%%%%%%%%%%%%%%%%%%%%%%%%%%%%%%%%%%%%%%%%%%%%%%%%%%%%%%%%%%%%
\label{s:LB}

\begin{theorem}
\label{thm:LB}
Let $Q$ be a non-uniform p.m.f.\ of full support on $A$.
Then the set $B_n$
in~\eqref{eq:Bn} satisfies,
\begin{equation*}
\big|\big\{x^n: H(\hat{P}_{x^n})\leq H(Q)\big\}\big| 
\gtrsim_{Q,m} 2^{nH(Q)}n^{\frac{(m-3)}{2}},
\end{equation*}
for all $n\geq N(Q,m)$, for an integer
$N(Q,m)$ that depends on $Q$ and $m$ only.
\end{theorem}

A key element of the proof of Theorem~\ref{thm:LB} is the following
estimate.

\begin{lemma} \label{lemmalattice}
Under the assumptions 
of Theorem~\ref{thm:LB}, we have
\begin{equation*}
    \Big|\Big\{P \in \mathcal{P}_n:  H(P) 
\in \Big[H(Q)-\frac{1}{n},H(Q)\Big]\Big\}\Big| \gtrsim_{Q,m} n^{m-2}+ n^{m-3},
\end{equation*}
for all $n\geq N'(Q,m)$, for an integer
$N'(Q,m)$ that depends on $Q$ and $m$ only.
\end{lemma}

\noindent
{\sc Proof. }
Let $d = m-1$.
We identify each $y$ in the simplex 
$$\mathcal{S} = \Big\{z \in \R^d: \sum_{i=1}^d z_i \leq 1, z_i \geq 
0\Big\},$$
with a p.m.f.\ $P_y \in \mathcal{P}$ by setting 
$P_y = (y,1-\sum_{i=1}^d y_i) \in \R^m$. We define the 
following sets in $\R^d$: 
\begin{align*}
    \clA_{Q,n} &= \Big\{y \in \mathcal{S}: H(P_y) > H(Q) - \frac{1}{n}\Big\}\\
    \clA_{Q}  &= \{y \in \mathcal{S}: H(P_y) > H(Q) \}.
\end{align*}

The set of all $n$-types $P \in \mathcal{P}_n$ such 
that $H(P)\in [H(Q)-1/n,H(Q)]$ has the same 
cardinality
(up to constants, since we define $\clA_{Q,n}$ with $>$ rather 
than $\geq$ in order for it to be open) as the set of lattice points 
\begin{align}
\Big(\frac{1}{n}\cdot\mathbb{Z}^d\Big) 
\cap (\clA_{Q,n} \cap \clA_Q^c) 
&= \Big(\frac{1}{n}\cdot\mathbb{Z}^d\Big) \cap \clA_{Q,n} 
- \Big(\frac{1}{n}\cdot\mathbb{Z}^d\Big) \cap \clA_{Q} \nonumber\\ 
\label{latticediff}
&= \mathbb{Z}^d\cap \bigl({n}\cdot \clA_{Q,n}\bigr) 
- \mathbb{Z}^d \cap \bigl({n}\cdot \clA_{Q}\bigr).
\end{align}

Note that by the concavity of entropy, 
both $\clA_Q$ and $\clA_{Q,n}$ are convex bodies. 
Our goal is to use Lemma~\ref{taovu} with 
$\Gamma = \mathbb{Z}^d$, generated by
$\{v_1,\ldots,v_d\}=\{e_1,\ldots,e_d\}$
being the standard basis
in $\RL^d$, and with $R=n$, in order 
to estimate $\big|\mathbb{Z}^d\cap \bigl({n}\cdot \clA_{Q,n}\bigr)\big|$ 
and $\big|\mathbb{Z}^d \cap \bigl({n}\cdot \clA_{Q}\bigr)\big|$ separately. 

For the latter, Lemma~\ref{taovu} applies directly with $B = \clA_Q$ to give 
\begin{equation} \label{firstlattice}
    \big|\mathbb{Z}^d \cap \bigl({n}\cdot \clA_{Q}\bigr)\big| 
= \Big(n^d + O_{Q,d}(n^{d-1})\Big)\mathrm{Vol}(\clA_Q).
\end{equation}

Since $Q$ has full support and is not the uniform distribution, the 
gradient of the entropy functional, $y \mapsto H(P_y)$, $y \in \mathcal{S}$
is bounded above near the boundary of $\clA_Q$, and thus we may apply 
a multivariate, first-order Taylor expansion to see that, for $n$ large 
enough depending on $Q$, and for some appropriately small constant $K(Q)$ 
that depends on $Q$ only,
\begin{equation} \label{ballinclusion}
\clA_Q+B_2\Bigl(\frac{K(Q)}{n}\Bigr) \subset \clA_{Q,n}.
\end{equation}
Therefore,
\begin{equation}
    \big|\mathbb{Z}^d \cap \bigl(n \cdot \clA_{Q,n}\bigr)\big| \geq  
\Big|\mathbb{Z}^d \cap 
\Bigl[n \cdot \Big(\clA_Q+B_2\Big(\frac{K(Q)}{n}\Big)\Big)\Bigr]\Big|.
\end{equation}
Now we claim that 
\begin{equation} \label{claim}
    \Big|\mathbb{Z}^d \cap \Bigl[n \cdot 
	\Big(\clA_Q+B_2\Big(\frac{K(Q)}{n}\Big)\Big)\Bigr]\Big| 
	\geq \Bigl(n^d + O_{Q,d}(n^{d-1})\Bigr)
	\mathrm{Vol}\Bigl(\clA_Q+B_2\Big(\frac{K(Q)}{n}\Big)\Bigr).
\end{equation}
Lemma~\ref{taovu} does not apply directly, since the error term 
depends on the set $B$ and we wish to 
take $B= \clA_Q+B_2(K(Q)/n)$, which depends 
on $n$. Nevertheless, examining the proof in~\cite{TV06:book}, we see 
that that the error term is -- up to constants depending 
only on the dimension -- 
the volume of the surface 
of $R\cdot B = n\cdot \bigl(\clA_Q+B_2(K(Q)/n) \bigr)$, which is 
still of the same order, i.e., 
\begin{equation} \label{replaceSur}
    \mathrm{Sur}\left(n \cdot \Big(\clA_Q+B_2\Big(\frac{K(Q)}{n}\Big)\Big)\right) 
= O_{Q,d}{(n^{d-1})}.
\end{equation}
Thus, repeating the proof of 
Lemma~\ref{taovu} as in~\cite{TV06:book}, but replacing the estimate 
of the error term 
with~\eqref{replaceSur}, we conclude that~\eqref{claim} holds.

Combining~\eqref{latticediff} with the 
estimates~\eqref{firstlattice} and~\eqref{claim}, we conclude that 
\begin{align} \nonumber
     &\Big|\Big\{P\in \mathcal{P}_n: H(P) \in \
	\Big[H(Q)-\frac{1}{n},H(Q)\Big]\Big\}\Big|  \\
     &\geq  \big(n^d+ O_{Q,d}(n^{d-1})\big)
\Bigl[\mathrm{Vol}\Bigl(\clA_Q+B_2\Big(\frac{K(Q)}{n}\Big)\Bigr) 
-\mathrm{Vol}(\clA_Q)\Bigr]  \\ \label{tworedstars}
     &\gtrsim_{Q,d} n^{d-1} + n^{d-2}
\end{align}
for $n$ large enough depending on $Q$ and $d$, where the implied constant in the last inequality depends to first order on $K(Q)$ and the 
surface area of $A_Q$. 

The result follows upon replacing $d$ by $m-1$.
\qed

\noindent
{\sc Proof of Theorem~\ref{thm:LB}. }
By Lemma~\ref{lem:stirling},
for any $n$-type $P$ with full support and 
$H(Q)-\frac{1}{n}\leq H(P) \leq H(Q)$,
\begin{equation} \label{typeclasssize}
    |\mathcal{T}(P)| \gtrsim_{Q,m} n^{-\frac{m-1}{2}}2^{nH(Q)}.
\end{equation}
Therefore, 
\begin{align} \nonumber
    &|\{x^n: H(\hat{P}_{x^n})\leq H(Q)\}| \\ \nonumber
	&\geq \sum_{P: H(P) \leq H(Q)}|\mathcal{T}(P)| \\
	\nonumber
    &\geq \sum_{P: H(Q)-\frac{1}{n}\leq H(P) \leq H(Q)}|\mathcal{T}(P)| \\ \label{lastcounting}
    &\gtrsim_{Q,m} n^{-\frac{m-1}{2}}2^{nH(Q)}
	\Big|\Big\{P \in \mathcal{P}_n:  H(P) \in \Big[H(Q)-\frac{1}{n},
	H(Q)\Big]\Big\}\Big|,
\end{align}
where in the last step we used~\eqref{typeclasssize}.

Combining~\eqref{lastcounting} with Lemma~\ref{lemmalattice} we obtain 
that, for $n$ large enough depending on $Q$ and $m$,
\begin{equation*}
    |\{x^n: H(\hat{P}_{x^n})\leq H(Q)\}| \gtrsim_{Q,m} n^{m-2} \frac{1}{n^{\frac{m-1}{2}}}2^{nH(Q)} = n^{\frac{m-3}{2}}2^{nH(Q)},
\end{equation*}
which is the claimed result. 
\qed

\subsection{Upper bound}
%%%%%%%%%%%%%%%%%%%%%%%%%%%%%%%%%%%%%%%%%%%%%%%%%
\label{s:UB}

Next we establish a matching upper bound.
In fact, we prove a stronger result, where the 
polynomial prefactor, the implied constants,
and the value of $n$ beyond which the bound holds
are the 
same for all $Q'$ a neighbourhood of $Q$. 

\begin{theorem}
\label{thm:UB}
Let $Q$ be a non-uniform p.m.f.\ of full support on $A$.
Suppose $\{Q_n\}_{n\geq1}$ be a 
sequence of p.m.f.s on $A$, such that $Q_n \to Q$ as $n \to \infty$. 
Then
\begin{equation*}
% \label{eq:mainLG}
\big|\{x^n\in A^n:\ H(\hat{P}_{x^n})\le H(Q_n)\}\ 
\big| \lesssim_{Q,m}\ n^{\frac{m-3}{2}}\,2^{nH(Q_n)},
\end{equation*}
for all $n\geq n_0(Q,m)$,
where $n_0(Q,m)$ depends only on $Q$ and $m$.
\end{theorem}

\noindent
{\sc Proof. }
Assume $n$ is large enough such that $Q_n$ is not uniform and has full support. 
We have 
\begin{equation} \label{firstsum}
\big|\{x^n\in A^n:\ H(\hat{P}_{x^n})\le H(Q_n)\}\big| 
= \sum_{r=1}^m\sum_{j\geq 0}\sum_{P \in \clS_{j,r}}|\mathcal{T}(P)|,
\end{equation}
where, for $r\in\{1,\dots,m\}$ and $j \in \mathbb{Z}_+$,  
\[
\clS_{j,r}:=\Bigl\{P\in\clP_n:\ |\mathrm{supp}(P)|=r,\ \frac{j}{n}\leq H(Q_n)-H(P) < \frac{j+1}{n}\Bigr\}.
\]

Applying Lemma~\ref{lem:stirling} and noting that on $\clS_{j,r}$ we have $2^{nH(P)}\le 2^{nH(Q_n)}\,2^{-j}$, we see that
\begin{equation}\label{eq:basic-slab}
\sum_{P\in\clS_{j,r}} |\clT(P)|
\ \lesssim_r \frac{2^{nH(Q_n)}\,2^{-j}}{n^{(r-1)/2}}
\ \sum_{P\in\clS_{j,r}}\ \prod_{a\in \mathrm{supp}(P)} \frac{1}{\sqrt{P(a)}}.
\end{equation}
Setting
\begin{equation}\label{eq:defJn}
J_n:=\Big\lceil (m+3)\,\log_2 n \Big\rceil,
\end{equation}
we split the inner sum in~\eqref{firstsum} as
\begin{equation} \label{splitsum}
\sum_{j\le J_n}\ \sum_{P\in\clS_{j,r}} |\clT(P)|\quad+\quad
\sum_{j>J_n}\ \sum_{P\in\clS_{j,r}} |\clT(P)|.
\end{equation}

First we show that second term in~\eqref{splitsum} has negligible contribution due to the exponential factor. 
For a fixed $r$ and $j > J_n$ we have
\[
|\clS_{j,r}| \le\ |\{P \in \clP_n: |\mathrm{supp}(P)| = r\}| \lesssim n^{r-1},
\]
and for any type $P$ with $r$ positive coordinates,
\[\prod_{a\in\mathrm{supp}(P)}\tfrac{1}{\sqrt{P(a)}}\le 
(\sqrt{n})^{\,r}=n^{r/2},
\]
since each $P(a)\ge 1/n$ whenever it is positive.
Substituting these into~\eqref{eq:basic-slab} yields
\[
\sum_{P\in\clS_{j,r}} |\clT(P)|
\ \lesssim_m \,2^{nH(Q_n)}\,2^{-j}\, n^{\,r-\frac12},
\]
where we noted that the
dependence of the implied constant on $r$ is equivalent to dependence on $m$.
Hence,
\begin{align*}
\sum_{j>J_n}\ \sum_{P\in\clS_{j,r}} |\clT(P)|
\ &\lesssim_{Q,m} 2^{nH(Q_n)}\,n^{\,r-\frac12}\sum_{j>J_n}2^{-j} \\
\ &\lesssim_{Q,m} 2^{nH(Q_n)}\,n^{\,r-\frac12}\,2^{-J_n} \\
% \ &\leq 2^{nH(Q_n)}\,n^{\,r-\frac12-(m+3)} \\
\ & 
% \leq 2^{nH(Q_n)}n^{-\frac{7}{2}} = 
=o_{Q,m}\big(2^{nH(Q_n)}\,n^{\frac{m-3}{2}}\big),
\end{align*}
where we used~\eqref{eq:defJn} and 
the fact that $r\le m$. Summing over $r$ we conclude that
\begin{equation} \label{FAR}
    \sum_{r=1}^m\sum_{j > J_n}^{n-1}\sum_{P \in \clS_{j,r}}|\mathcal{T}(P)| = o_{Q,m}\big(2^{nH(Q_n)}\,n^{\frac{m-3}{2}}\big).
\end{equation}

We now bound the first term in~\eqref{splitsum}, 
corresponding to $j\le J_n$. Let $\delta = \delta(Q)>0$ to 
be chosen later. 
For notational convenience, for the rest of the proof
we take, without loss of generality, the alphabet $A$ to be
$\{1,2,\ldots,m\}$.
For $1\leq r\leq m$ and $s=0,1,\dots,r$, let
\[
\clS_{j,r}^{(s)}:=\Big\{P\in\clS_{j,r}: |\{i:\ P(i)\le \delta\}|=s\Big\},
\]
and write
\[
F(P):=\{i:\ P(i)\le \delta\},\quad\mbox{so that}\;|F(P)|=s
\;\mbox{for}\;
P \in S_{j,r}^{(s)}.
\]
We bound $\sum_{P\in\clS_{j,r}}|\clT(P)|$ by splitting over $s$ and $F$. 
Clearly, since $j\leq J_n$ and $J_n/n\to 0$, by choosing $\delta$ small 
enough depending only on $Q$ and taking $n$ large enough so 
that $J_n/n$ is small and $H(Q_n)$ is close to $H(Q)$, there have 
to be at least two coordinates of any $P \in \clS_{j,r}^{(s)}$ 
with $P(i) > \delta$ (as, otherwise, $H(P)$ would be smaller 
than $H(Q_n) - \epsilon$ for some $\epsilon > 0$). Therefore, 
it suffices to restrict to $s\leq r - 2$:
\begin{equation*}
\sum_{P\in\clS_{j,r}}|\clT(P)|
= \sum_{s=0}^{r-2}\ \sum_{\substack{F\subseteq A\\ |F|=s}}\ \sum_{P\in\clS_{j,r}^{(s)}:\ F(P)=F} |\clT(P)|. 
% \label{eq:split-s}
\end{equation*}
By Lemma~\ref{lem:stirling} and 
the fact that $2^{nH(P)}\le 2^{nH(Q_n)}\,2^{-j}$ on $\clS_{j,r}$,
\[
|\clT(P)|\ \lesssim_{m} \frac{2^{nH(Q_n)}\,2^{-j}}{n^{(r-1)/2}}\ \prod_{i=1}^r \frac{1}{\sqrt{P(i)}}.
\]
Hence,
\begin{align}
\sum_{P\in\clS_{j,r}^{(s)}:\ F(P)=F} |\clT(P)|
&\lesssim_{m} \frac{2^{nH(Q_n)}\,2^{-j}}{n^{(r-1)/2}}
\sum_{P\in\clS_{j,r}^{(s)}:\ F(P)=F}
\Big(\prod_{i\notin F}\frac{1}{\sqrt{P(i)}}\Big)\Big(\prod_{i\in F}\frac{1}{\sqrt{P(i)}}\Big) \nonumber\\
&\lesssim_{m,\delta} \frac{2^{nH(Q_n)}\,2^{-j}}{n^{(r-1)/2}}
\sum_{P\in\clS_{j,r}^{(s)}:\ F(P)=F}
\ \prod_{i\in F}\frac{1}{\sqrt{P(i)}},
\label{eq:large-small}
\end{align}
where we used $P(i)\ge\delta$ for $i\notin F$, so $\prod_{i\notin F}P(i)^{-1/2}\le \delta^{-(r-s)/2}$.

Now parametrise the $s$ ``small'' coordinates by their 
counts $c_i:=nP(i)\in\{1,2,\dots,\lfloor \delta n\rfloor\}$ 
(note that, as before, $c_i\ge1$ whenever $P(i)>0$).
After fixing these counts, the remaining $r-s$ coordinates then live on a compact subset of $\R^{r-s-1}$. Moreover, the entropy of the type has to be 
concentrated on those coordinates, as the entropy contribution of the rest can be made arbitrarily small by choosing $\delta(Q)$ small enough. Thus, we can use an analogous argument to~\eqref{ballinclusion}--\eqref{tworedstars}, but with the corresponding upper bounds, to conclude that for $n$ large enough depending on $Q$ and $m$ there are 
\begin{equation} \label{countasbefore}
    \lesssim_{Q,m} n^{\,r-s-2} \quad \text{ lattice points (uniformly for $j\le J_n$) corresponding to $P \in \clS_{j,r}^{(s)}$.} 
\end{equation}
Note that $\clA_{Q_n}$ is still a convex body and 
its boundary is a smooth hypersurface, so we still have
$$
\mathrm{Sur}\left(n \cdot \Big(\clA_{Q_n}+B_2\Big(\frac{K(Q)}{n}
\Big)\Big)\right) 
= O_{Q,d}{(n^{d-1})}.
$$
[In fact, the implied constant in this estimate can be taken to be 
independent of $Q$, since $\clA_{Q_n}$ is a convex subset 
of the simplex]. The reverse inclusion to~\eqref{ballinclusion} can be 
justified by observing that the entropy is strictly concave,
and since $Q$ is not the uniform distribution, the gradient is 
also uniformly bounded below coordinatewise near the boundary 
of $\clA_{Q_n}$. The analogous upper bound to~\eqref{tworedstars} 
can also be justified in the same way as~\eqref{tworedstars},
since $\mathrm{Vol}(C + B_2(\epsilon))$ is locally polynomial 
in $\epsilon$ for any convex body $C$.

Therefore, by~\eqref{countasbefore}, for $n$ large enough depending on $Q$ and $m,$
\begin{align}
\sum_{P\in\clS_{j,r}^{(s)}:\ F(P)=F}
\ \prod_{i\in F}\frac{1}{\sqrt{P(i)}}
&\lesssim_{Q,m} \sum_{(c_i)_{i\in F}}\ \Big(\prod_{i\in F}\sqrt{\frac{n}{c_i}}\Big)\ \cdot n^{\,r-s-2} \nonumber\\
&\le \Big(\prod_{i\in F}\sum_{c=1}^{\lfloor \delta n\rfloor} \sqrt{\frac{n}{c}}\Big)\ \cdot n^{\,r-s-2}
\nonumber\\
&\lesssim_{\delta,m}n^{\,s}\, n^{\,m-s-2},
\label{eq:sum-small}
\end{align}
since we obviously have
$$\sum_{c=1}^{\lfloor\delta n\rfloor}\sqrt{n/c}\le \sqrt{n}\sum_{c=1}^{\lfloor\delta n\rfloor}c^{-1/2}\lesssim_{\delta}n.$$

Combining~\eqref{eq:large-small} and~\eqref{eq:sum-small}, and summing over 
all $\binom{r}{s}$ choices for $F$,
we get, for $n$ large enough depending on $Q,m,$
\begin{align*}
\sum_{P\in\clS_{j,r}^{(s)}} |\clT(P)|
&\lesssim_{Q,m} 2^{nH(Q_n)}\,2^{-j}{n^{-(r-1)/2}}\ 
\cdot\ n^{\,r-s-2}\, n^{\,s}\\
&=2^{nH(Q_n)}\,2^{-j}\, n^{\frac{r-3}{2}}\\
&\leq 2^{nH(Q_n)}\, 2^{-j}\,n^{\frac{m-3}{2}},
\end{align*}
since $r\leq m$.
Summing over $1\leq r\leq m$, $0\leq s\leq m-2$, and $j\le J_n$ yields:
\begin{equation} \label{NEAR}
\sum_{r=1}^m\sum_{j\le J_n}\ \sum_{P\in\clS_{j,m}} |\clT(P)|
\lesssim_{Q,m} 2^{nH(Q_n)}\, n^{\frac{m-3}{2}}.
\end{equation}

Finally, combining~\eqref{FAR} and~\eqref{NEAR} yields the claimed bound. 
\qed

\section{Universal compression}
%%%%%%%%%%%%%%%%%%%%%%%%%%%%%%%%%%%%%%%%%%%%%%%%%%%%%%%%%%%%%%%%%%%%%%
\label{s:universal}

\begin{theorem}[Universal achievability]
\label{thm:universal}
Consider all memoryless sources $\Xp$ on a finite alphabet
$A$ of size $|A|=m\geq 2$.
There exists a sequence of universal,
one-to-one, variable-rate 
compressors $\{\phi_n^*\}$ such that,
on any source with distribution $P$ with full support on $A$,
and for any $\delta \in (0, D(U\|  P))$,
the compressors $\{\phi_n^*\}$ 
simultaneously achieve excess-rate probability,
\begin{equation}
\BBP(\ell(\phi_n^*(X^n))\geq nR^{\sf u}(n,\delta,P)\big)
\leq 2^{-n\delta}
\quad\mbox{eventually},
\label{eq_goal_error}
\end{equation}
at a rate $R^{\sf u}(n,\delta,P)$ that satisfies,
as $n\to\infty$,
\begin{equation}
nR^{\sf u}(n,\delta,P)\leq n H(P_{\alpha^*}) 
+ \Big(\frac{m-2}{2} - \frac{1}{2(1-\alpha^* )} \Big) \log n + O(1),
\label{eq_goal_rate}
\end{equation}
where $\alpha^*$ is the unique $\alpha \in (0,1)$ for 
which $\delta=D(P_{\alpha^*}\|  P)$,
the p.m.f.\ $P_\alpha$ is defined in~{\em (\ref{eq:Palpha})},
and the implied constant in the $O(1)$ term
depends on $m,P$ and $\delta$.
\end{theorem}

\noindent
{\bf Remark. }
An examination of the proof below shows that 
the excess-rate probability bound in~(\ref{eq_goal_error}) holds
for all $n$ greater than some 
$M_0(P,\delta)$, which can be explicitly identified
in terms of $P$ and $\delta$ in a similar way
to how the expression for $N_0$ was derived
in Theorem~\ref{thm_converse}. But in the 
present context, the exact value of $M_0(P,\delta)$
is less relevant, since our bound
on the rate in~(\ref{eq_goal_rate}) is
already an asymptotic statement.

\medskip

The observation in the following lemma,
which follows directly from the definitions,
will be useful in the proof of Theorem~\ref{thm:universal}.

\begin{lemma}
\label{lem_technical_D_and_H}
For all $\alpha \in (0,1)$, and any pair of p.m.f.s $P,Q$ 
on $A$ such that $P$ has full support,
we have:
    \begin{equation*}
        \alpha [ D(Q \|P) - D( P_\alpha \|P)   ] 
= D(Q \| P_\alpha) + (1-\alpha)[ H(Q)-H(P_\alpha)].
    \end{equation*}
\end{lemma}

\noindent
{\sc Proof of Theorem~\ref{thm:universal}. }
The construction of the compressors
$\{\phi_n^*\}$ is very simple.
For each $n$, all strings $x^n\in A^n$ 
are sorted in
order of increasing empirical entropy
$H(\hat{P}_{x^n})$, with ties broken
arbitrarily. This ordering is shared
between the encoder and decoder.
Then, each $x^n$ is described by
the binary representation of its index
in the ordered list, so that
the $k$th string $x^n$ has a description
length of 
$\ell(\phi_n^*(x^n))=\lfloor \log k\rfloor$ bits.

Choose and fix a p.m.f.\ with full support on $A$
and a $\delta\in(0,D(U\|P))$. As in Theorems~\ref{thm_achievability}
and~\ref{thm_converse}, let $\alpha^*$ be the 
unique $\alpha\in(0,1)$ such that $D(P_{\alpha^*}\|P)=\delta$.
For each $n\geq 1$, let $\alpha_n\in(0,1)$ be 
a constant to be chosen later. Consider
the sets
\begin{equation*}
    E_n = 
	\{   x^n \in A^n :
    H(\hat{P}_{x^n}) \leq H(P_{\alpha_n}) \},
\end{equation*}
and let $R^{\sf u}(n,\delta,P) = \frac{1}{n}[\log |E_n|+1]$.

We first bound the excess-rate probability of $\phi_n^*$.
This step will take up most of the
proof. Specifically, we show that there is
a sequence $\{\alpha_n\}$ such that,
for all $n$ large enough, $\alpha_n\in[\alpha^*,1)$ and
\begin{equation*}
\BBP\big(\ell(\phi_n^*(X^n))\geq nR^{\sf u}(n,\delta,P)\big)=
\BBP\big(\ell(\phi_n^*(X^n))\geq\log |E_n|+1\big)\leq P^n(E_n^c)\leq 2^{-n\delta},
\end{equation*}
which implies~(\ref{eq_goal_error}).
The equality and the first inequality follow from the
definitions, so we need to show that
\bqq
P^n(E_n^c)\leq 2^{-n\delta}\quad\mbox{eventually}.
\label{eq:targetP}
\eqq
To that end, we begin by expressing
$$P^n (E_n^c) 
= 
	\sum_{x^n \in A^n} P^n(x^n) \mathbb{I}_{E_n^c}(x^n)  
=  
	\sum_{x^n \in A^n} P_{\alpha^*}^n(x^n) \exp 
	\Big(\log_e \frac{P^n(x^n)}{P_{\alpha^*}^n(x^n)} \Big)
	\mathbb{I}\{ H(\hat{P}_{x^n}) > H(P_{\alpha_n}) \}.
$$
Noting that, for every $x^n$,
$$H(\hat{P}_{x^n})=\frac{1}{n}\sum_{i=1}^n\log\frac{1}{P(x_i)}
-D(\hat{P}_{x^n}\|P),$$
we can further express $P^n(E_n^c)$ as
\begin{align*}
P^n (E_n^c) 
&=  
	\sum_{x^n \in A^n} P_{\alpha^*}^n(x^n) \exp 
	\Big(\log_e \frac{P^n(x^n)}{P_{\alpha^*}^n(x^n)} \Big) \nonumber \\
& \hspace{0.8in}
	\mathbb{I} \left\{ H(\hat{P}_{x^n}) > H(P_{\alpha_n}), 
	\;\frac{1}{n} \sum_{i=1}^n \log \frac{1}{P(x_i)}> H(P_{\alpha_n}) 
	+ D(\hat{P}_{x^n}\|P)  \right\}  \\
&\leq  
	\sum_{x^n \in A^n} P_{\alpha^*}^n(x^n) \exp 
	\Big( - \sum_{i=1}^n \log_e \frac{P_{\alpha^*}(x_i)}{P(x_i)} \Big)
	\mathbb{I}\Bigg\{ \frac{1}{n} \sum_{i=1}^n \log \frac{1}{P(x_i)}> 
	H(P_{\alpha_n}) + D(P_{\alpha_n}\|P) \Bigg\}, 
\end{align*}
where the last inequality follows from Lemma~\ref{lem_technical_D_and_H}.
Recalling the definition of $Z_\alpha$ from~(\ref{eq:Palpha})
and defining $\beta_n>0$ via
$H(P_{\alpha_n}) + D(P_{\alpha_n}\|P)=-\frac{\log \beta_n}{n}$, we 
have
\begin{align*}
P^n (E_n^c) 
&\leq  
	\sum_{x^n \in A^n}\Bigg[ P_{\alpha^*}^n(x^n) \exp 
	\Big( - \sum_{i=1}^n \log_e \frac{P_{\alpha^*}(x_i)}{P(x_i)} \Big)\\
&\hspace{1in} 
	\mathbb{I}  \Bigg\{      n\log_e Z_{\alpha^*} + \sum_{i=1}^n \log_e  
	P^{1-\alpha^*}(x_i) < n\log_e Z_{\alpha^*} 
	+(1-\alpha^*) \log_e\beta_n\Bigg\}\Bigg]\\
&=  
	\mathbb{E}_{P_{\alpha^*}^n} \Bigg[ \exp \Big( - \sum_{i=1}^n \log_e 
	\frac{P_{\alpha^*}(X_i)}{P(X_i)} \Big)\\
& \hspace{1in} 
	\mathbb{I} \Bigg\{\sum_{i=1}^n \log_e 
	\frac{P_{\alpha^*}(X_i)}{P(X_i)} > - n\log_e Z_{\alpha^*} -(1-\alpha^*) 
	\log_e \beta_n \Bigg\}  \Bigg].  
\end{align*}
We further bound this expression
using the Berry-Ess\'{e}en bound
in the form given in Lemma~\ref{lem BE exponential},
with $Z_i= \log_e [P_{\alpha^*}(X_i)/P(X_i)]$ 
and $x=- n\log_e Z_{\alpha^*} -(1-\alpha^*) \log_e \beta_n$,
to obtain that
\begin{align}
    P^n (E_n^c)  & \leq \frac{1}{\sigma_{2}} \Big(  \frac{1}{\sqrt{2\pi}} + \frac{\rho_{2}}{\sigma_{2}^2} \Big) \frac{1}{\sqrt{n}} \exp \Big( n\log_e Z_{\alpha^*} +(1-\alpha^* ) \log_e \beta_n   \Big)    
	\nonumber\\ 
    &=  \frac{1}{\sigma_{2}} \Big( \frac{1}{\sqrt{2\pi}} + \frac{\rho_{2}}{\sigma_{2}^2}  \Big) \frac{Z_{\alpha^*}^n \beta_n^{1-\alpha^*}}{ \sqrt{n}},
\label{err_prob2}
\end{align} 
where $\sigma_2$ and $\rho_2$ are as in
Theorem~\ref{thm_achievability}.
Let
$\bar{r}= \frac{1}{1-\alpha^*} \log 
\big(\frac{1}{\sigma_{2}} (\frac{1}{\sqrt{2\pi}} 
+ \frac{\rho_{2}}{\sigma_{2}^2})\big)$.
If $\beta_n$ satisfies
\begin{align}
\log \beta_n 
&
	\leq \frac{1}{1-\alpha^*} \Bigg[ -n \delta - \log 
	\Bigg(  \frac{1}
	{
		\sigma_{2}} \Big(\frac{1}{\sqrt{2\pi}} 
		+ \frac{\rho_{2}}{\sigma_{2}^2} \Big)\frac{Z_{\alpha^*}^n }
		{\sqrt{n}}
	\Bigg) \Bigg] \nonumber \\
&= 
	-\frac{1}{1-\alpha^*}(\delta + \log Z_{\alpha^*})n 
      +\frac{1}{2(1-\alpha^*)} \log n -\bar{r}, 
      \label{eq_goal_on_beta_n}
\end{align}
then~(\ref{err_prob2})
immediately implies that~(\ref{eq:targetP}) holds,
which gives~\eqref{eq_goal_error}, as required.

\newpage

So it only remains to show that $\{\alpha_n\}$ can be selected
so that~\eqref{eq_goal_on_beta_n} holds.
In order to choose $\{\alpha_n\}$ appropriately,
we recall that $\alpha_n\geq\alpha^*$ eventually
and that the Taylor expansion~(\ref{eq taylor D 2})
in the proof of Theorem~\ref{thm_converse}
gives,
for all $n$ large enough,
\begin{equation}
    D(P_{\alpha_n}\|  P) 
\geq \delta - (\alpha_n - \alpha^*) (1- \alpha^* ) \sigma_3 ^2 (\log e) - \frac{\log e}{2} (\alpha_n - \alpha^*)^2 \Big| \tilde{\sigma}_3^2 
-(1-\alpha^*)\hat{\rho}_3\Big|,
\label{eq_D_1}
\end{equation}
where the constants $\sigma_3$, $\tilde{\sigma}_3$ 
and $\hat{\rho}_3$ are defined in 
Lemma~\ref{lem var, 3rd mom} and
Theorem~\ref{thm_converse}.
Similarly, the Taylor expansion~(\ref{eq taylor H 2})
in the proof of Theorem~\ref{thm_converse}
gives,
for all $n$ large enough,
\begin{equation}
    H(P_{\alpha_n})
\geq H(P_{\alpha^*})  - (\alpha_n - \alpha^*) \alpha^* \sigma_3 ^2  \log e  - \frac{\log e}{2} (\alpha_n - \alpha^*)^2 (\hat{\sigma}_3 ^2 + 
\hat{\rho}_3)  
\label{taylor_H1},
\end{equation}
where the constant $\hat{\sigma}_3$
is defined in 
Theorem~\ref{thm_converse}.
Therefore, combining~(\ref{eq_D_1})
with~(\ref{taylor_H1}), and recalling that 
$\delta=D(P_{\alpha^*} \| P)$, we have
\begin{align}
     H(P_{\alpha_n})+D(P_{\alpha_n}\|  P) &\geq H(P_{\alpha^*})+ 
	\delta 
	- (\alpha_n - \alpha^*)\bar{p}
	- (\alpha_n - \alpha^*)^2 \bar{q},
	 \label{eq:target1}
\end{align}
where $\bar{p}=\sigma_3^2(\log e) >0$ 
and $\bar{q}=\frac{\log e}{2}\big(|\tilde{\sigma}_3^2-(1-\alpha^*)
\hat{\rho}_3| + \hat{\sigma}_3 ^2 + \hat{\rho}_3 \big) >0$.

Now we are in a position to choose $\{\alpha_n\}$. Letting,
for each $n$,
\begin{equation*}
    \alpha_n = \alpha^* + \frac{1}{2\bar{p}(1-\alpha^*)}  
\frac{\log n}{n} 
	- \frac{\bar{q}+\bar{r}}{\bar{p}}  \frac{1}{n},  
\end{equation*}
we see that $1>\alpha_n\geq \alpha^*$ eventually.
Also, simple algebra shows that, eventually,
$$(\alpha_n - \alpha^*)\bar{p}
+ (\alpha_n - \alpha^*)^2 \bar{q}
\leq  \frac{1}{2(1-\alpha^*)} 
\frac{\log n}{n} -\frac{\bar{r}}{n}, \label{ineq_for_alpha_n}$$
which, recalling from~(\ref{eq:PalphastarZ}) 
that $H(P_{\alpha^*}) 
=\frac{1}{1-\alpha^*}(\alpha^*\delta + \log Z_{\alpha^*})$,
gives,
\bqq
H(P_{\alpha^*})+ 
	\delta 
	- (\alpha_n - \alpha^*)\bar{p}
	- (\alpha_n - \alpha^*)^2 \bar{q} 
	\geq \frac{1}{1-\alpha^*}(\delta + \log Z_{\alpha^*}) 
	-\frac{1}{2(1-\alpha^*)} \frac{\log n}{n} +\frac{\bar{r}}{n}.
\label{eq:target2}
\eqq
Combining~(\ref{eq:target1}) with~(\ref{eq:target2})
and the definition of $\beta_n$ implies
that~\eqref{eq_goal_on_beta_n} holds,
which gives the required bound~(\ref{eq_goal_error}) 
on the excess-rate probability.

Finally, we will establish the claimed bound~(\ref{eq_goal_rate})
on the rate, $R^{\sf u}(n,\delta,P)=\frac{1}{n}[\log |E_n|+1]$. 
The key step in the evaluation of $R^{\sf u}(n,\delta,P)$
was already established in 
Theorem~\ref{thm:UB}, from which we obtain that,
as $n\to\infty$,
\bqq
\log |E_n|\leq nH(P_{\alpha_n})+\Big(\frac{m-3}{2}\Big)\log n
+O(1).
\label{eq:fromLG}
\eqq
Also, from the earlier Taylor expansion in~\eqref{eq taylor H 1}
we can bound, for $n$ large enough,
\begin{align}
    H(P_{\alpha_n})
    & \leq H(P_{\alpha^*})  - (\alpha_n - \alpha^*) \alpha^* \sigma_3 ^2  \log e  + \frac{\log e}{2} (\alpha_n - \alpha^*)^2 (\hat{\sigma}_3 ^2 +\hat{\rho}_3) \label{eq_taylor_H_UB}\\
    & \leq H(P_{\alpha^*}) - \Big(\frac{1}{2\bar{p}(1-\alpha^*)}  
	\frac{\log n}{n} 
	- \frac{\bar{q}+\bar{r}}{n} \Big) 
	\alpha^* \bar{p}  + \frac{\log e}{2n}  
	(\hat{\sigma}_3 ^2 +\hat{\rho}_3),
	\nonumber
\end{align}
where~\eqref{eq_taylor_H_UB} holds 
since $\sigma_{3,v}^2 \leq \hat{\sigma}_3^2$, $v \leq 1$, and 
by the definition of $\hat{\rho}_3$.
Therefore, 
as $n\to\infty$,
\bqq
H(P_{\alpha_n})
    \leq H(P_{\alpha^*})-\frac{\alpha^*}{2(1-\alpha^*)}  \frac{\log n}{n} 
+O\Big(\frac{1}{n}\Big),
\label{eq_taylor_H_UB_2}
\eqq
and combining~(\ref{eq:fromLG}) with~(\ref{eq_taylor_H_UB_2}) 
yields
the required bound~(\ref{eq_goal_rate})
on the rate and completes the proof.
\qed

\section{Concluding remarks}
%%%%%%%%%%%%%%%%%%%%%%%%%%%%%%%%%%%%%%%%%%%%%%%%%%%%%%%%%%%%%%%

This work revisits the problem of lossless data compression,
in the case when strict guarantees are required on the probability
that the target compression rate might be exceeded. When this
excess-rate probability is required to be very small, 
classical approximations to the best achievable 
target rate -- {\em \`{a}~la} Strassen -- fail to be informative.
Instead, much more appropriate approximations are obtained
by examining the best achievable rate when the excess-rate
probability is exponentially small in the blocklength.
The new approximations to the optimal
rate are derived from explicit nonasymptotic
bounds, that are accurate up to and including terms of order
$\frac{\log n}{n}$. These bounds support a crucial operational 
conclusion: In applications where the blocklength 
is only moderately large and where stringent guarantees are 
required on the rate, the best achievable rate is no longer 
close to the entropy. Instead, it is given by the more pragmatic rate
described by the inverse to the error exponent function.
These results apply to both one-to-one and prefix-free
variable rate codes.

In the case of universal compression, it is shown that the 
same performance can be universally achieved by one-to-one
codes, at a cost
of no more than $(\frac{m-2}{2})\log n$ bits, over the class 
of all memoryless
sources on an alphabet of size $m$. The key technical step
in the proof of this result is a precise evaluation of the
number of strings with empirical entropy below a certain
threshold.

There are several possible future directions in which 
the present results can be generalized and extended. 
Two immediate possibilities
are described by the two conjectures we formulated in
Section~\ref{s:Iuniversal}:

\begin{enumerate}
\item
The universal rate obtained in Theorem~\ref{thm:universal}
is optimal for one-to-one codes.
\item
The best universally achievable rate for prefix-free
codes incurs a penalty of $\frac{m}{2}\log n$ bits
over the optimal (non-universal) rate. In other words,
Theorem~\ref{thm:universal} remains valid for prefix-free
codes, and a matching converse also holds, 
with 
$\frac{m-2}{2}$ replaced by $\frac{m}{2}$
in the universal rate $R^{\sf u}(n,\delta,P)$.
\end{enumerate}

There are also several avenues for more ambitious
generalizations. For example, it would be interesting
to determine 
the corresponding {\em pragmatic} rate for the problem
of lossy data compression or, even more interestingly,
for channel coding. The results in this paper and the
earlier hypothesis testing results discussed in 
the introduction are encouraging indications that 
it might be possible to identify similar
phenomena in lossy data compression, in channel coding,
and perhaps in some of the many other 
scenarios that arise in multi-terminal communication
systems.

\bibliographystyle{plain}
%\bibliography{../../latex/ik}
\bibliography{ik.bib}

\end{document}